\documentclass[a4paper,11pt]{article}
\usepackage{jcappub}
\usepackage{orcidlink}
\usepackage[dvipsnames]{xcolor}

\usepackage{booktabs} 
\usepackage{multirow} 
\usepackage{rotating}

\usepackage{mathtools}
\usepackage{amsmath}
\usepackage{amsfonts}
\usepackage{amssymb}
\usepackage{textcomp}
\usepackage{bbold}
\usepackage{bm}
\usepackage{esint}
\usepackage{orcidlink}

\usepackage{soul}
\usepackage{slashed}

\newcommand{\bk}{{\bf k}}

\long\def\exclude#1{}

\title{Supernova Bounds on Neutrino Dipole Moments from a Modern Perspective}

\author[a]{Damiano F.\ G.\ Fiorillo \orcidlink{0000-0003-4927-9850},} 
\affiliation[a]{Gran Sasso Science Institute (GSSI), L’Aquila, Italy}

\author[b,c]{Martina La Rosa
\orcidlink{0009-0002-8500-3376},}
\affiliation[b]{Dipartimento di Fisica e Astronomia, Università degli Studi di Padova,
Via Marzolo 8, 35131 Padova, Italy}
\affiliation[c]{Istituto Nazionale di Fisica Nucleare (INFN), Sezione di Padova,
Via Marzolo 8, 35131 Padova, Italy}

\author[d]{Federico Testagrossa \orcidlink{0009-0000-9401-1971},}
\affiliation[d]{Deutsches Elektronen-Synchrotron DESY, Platanenallee 6, 15738 Zeuthen, Germany}

\author[b,c]{and Edoardo Vitagliano
\orcidlink{0000-0001-7847-1281}}

\abstract{
Neutrinos may have nonstandard magnetic dipole moments $\mu_\nu$. In the dense cores of supernovae (SNe), these would convert part of the neutrinos into sterile, right-handed states through the electromagnetic fluctuations of the charged plasma. The streaming sterile neutrinos with 100-MeV energies can convert back to standard neutrinos in the Galactic magnetic field, leading to a high-energy neutrino flux unobserved from the historical SN~1987A. We reassess the production rates, including scattering on both protons and electrons, and, building upon advances in SN simulations and in the description of the Milky Way magnetic field, we revisit the bounds on magnetic dipole moments. We find for Dirac electron neutrinos $\mu_{\nu_e} \lesssim 4.8 \times 10^{-13} \mu_B$, superseding all existing terrestrial and astrophysical constraints.
}
\begin{document}

\maketitle

\flushbottom
\section{Introduction\label{sec:intro}}
The study of neutrino electromagnetic interactions has a long history in particle physics and has attracted extensive efforts from astrophysics, cosmology, and laboratory~\cite{Raffelt:1996wa,Raffelt:1999gv,Giunti:2014ixa,Giunti:2024gec}. That neutrinos might possess a magnetic dipole moment (MDM) was hypothesized together with their very existence by Pauli in 1930~\cite{Pauli:1930pc}, and an experimental attempt to probe it was made shortly afterwards by Nahmias in 1935~\cite{Nahmias:1935yih}. The neutrino MDM received renewed interest when it was evoked as a solution to the solar neutrino puzzle~\cite{Cisneros:1970nq,Voloshin:1986ty,Okun:1986hi,Okun:1986na,Notzold:1988kz, Lim:1987tk, Akhmedov:1988uk, Akhmedov:1992en}. Although this explanation has faded away, the discovery of neutrino MDM would be one of the most intriguing indications of new physics. The Standard Model (SM) itself, when extended to account for neutrino masses, does predict finite, albeit small, MDMs through radiative corrections: nothing forbids the existence of much larger MDMs. Additionally, Majorana neutrinos cannot have flavor-diagonal MDMs, while Dirac neutrinos can. Thus, MDM detection would provide direct evidence for non-standard physics~\cite{Kim:1976gk,Beg:1977xz,Duncan:1987ki,Fukugita:1987ti,Babu:1987be} and insight into the fundamental properties of neutrinos, including the nature of their masses~\cite{Pal:1981rm,Shrock:1982sc,Schechter:1981hw,Nieves:1981zt,Kayser:1982br,Kayser:1984ge,Bell:2005kz,Bell:2006wi} and their role in astrophysical objects~\cite{Bernstein:1963qh,Fujikawa:1980yx,Raffelt:1989xu,Raffelt:1990pj,Raffelt:1992pi,Haft:1993jt,Viaux:2013hca,Arceo-Diaz:2015pva,Capozzi:2020cbu,Kopp:2022cug,Jana:2022tsa,Sasaki:2023sza,Ehrhardt:2026fis}.

Among such astrophysical environments, core-collapse supernovae (CCSNe) are quite special from the perspective of neutrino physics. Newly formed proto-neutron stars (PNSs) are so hot and dense that neutrinos, whose chemical potential can reach hundreds of MeV, are trapped and can finally stream outside only when they reach the so-called neutrinosphere, at the edge of
the SN core, where they leave with much smaller energies of tens of MeV. This emission lasts about ten seconds, after which the PNS temperature is small enough that even neutrinos produced at its center can freely stream out (for a list of reviews see e.g.~\cite{Janka:2012wk, Scholberg:2012id, Mirizzi:2015eza,Janka:2016fox,Vitagliano:2019yzm,Raffelt:2025wty}). This scenario was put to the test by the historical SN~1987A, with the detection of a neutrino signal at  Kamiokande-II~\cite{Kamiokande-II:1987idp,Hirata:1988ad,Hirata:1991td,Koshiba:1992yb} and Irvine–Michigan–Brookhaven (IMB)~\cite{Bionta:1987qt,IMB:1988suc} water Cherenkov experiments, together with the Baksan Underground Scintillation Telescope (BUST)~\cite{Alekseev:1987ej,Alekseev:1988gp}. These observations are broadly consistent with modern simulations of CCSNe~\cite{Fiorillo:2023frv}, and have offered a treasure trove for searches for new physics.

\begin{figure*}
    \centering
    \includegraphics[width=1.0\linewidth]{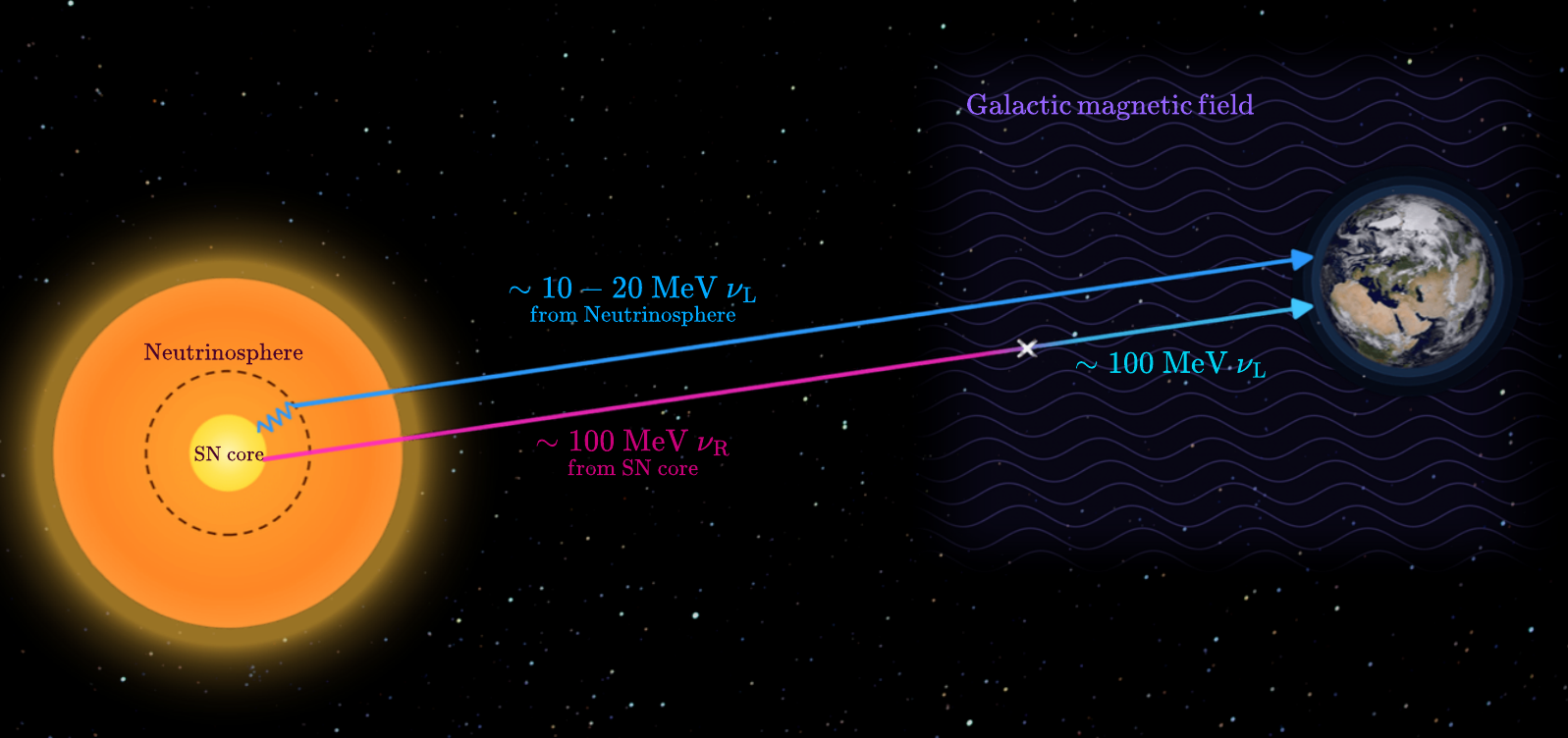}
    \caption{(Out-of-scale) Schematic representation of the two neutrino fluxes expected from SN~1987A. Deep in the PNS core, 100-MeV left-handed neutrinos can be converted into right-handed states by scattering on the electromagnetic field fluctuation sourced by the plasma, subsequently converting back into active states in the Galactic magnetic field, finally reaching IMB and Kamiokande-II. Such neutrinos will have cross sections much larger than the ones leaving from the neutrinosphere.
    }
    \label{neutrinos_from_sn}
\end{figure*}

If neutrinos are Dirac particles, their behavior in CCSNe would be drastically affected by the existence of MDMs. Their spins would be flipped by the electromagnetic fields sourced by protons and electrons, causing part of them to stream out of the core directly as right-handed, sterile states. The energies of these escaping particles are set by the hot core, and are thus at the level of $100\,\mathrm{MeV}$, much higher than the standard, trapped left-handed neutrinos diffusing away from the neutrinosphere with $10-20\,\mathrm{MeV}$ typical energies.
A schematic representation of this scenario is given in Fig.~\ref{neutrinos_from_sn}, showing how this mechanism results in a flux of left-handed neutrinos escaping the SN from the neutrinosphere and in a flux of  right-handed neutrinos leaving the core with $\sim 100 \ \mathrm{MeV}$ energies.

On the one hand, these high-energy particles offer a new energy-loss channel, resulting in the shortening of the expected neutrino signal~\cite{Raffelt:1996wa,Raffelt:2006cw,Caputo:2022rca,Fiorillo:2025yzf}. Therefore, similar to, e.g., axions~\cite{Raffelt:1990yz,Mayle:1987as,Mayle:1989yx,Burrows:1988ah,Fiorillo:2025gnd}, axion-like particles~\cite{Chang:2018rso,Croon:2020lrf,Camalich:2020wac,Caputo:2021rux,Ferreira:2022xlw,Fiorillo:2025sln}, and millicharged particles~\cite{Davidson:2000hf,Fiorillo:2024upk}, strong bounds on neutrino MDM come from the requirement that the additional cooling channel does not exceed the neutrino luminosity at 1 s after core bounce~\cite{Notzold:1988kz,Barbieri:1988nh,Ayala:1998qz,Kuznetsov:2007mp,Kuznetsov:2009zm}.

A much more stringent bound, proposed independently by Barbieri and Mohapatra~\cite{Barbieri:1988nh}, Lattimer and Cooperstein~\cite{Lattimer:1988mf}, and  N\"otzold~\cite{Notzold:1988kz} shortly after the SN~1987A event, obtains from the neutrino energy spectrum detected at Earth. Prompted by Galactic magnetic fields, right-handed neutrinos can indeed convert en route to Earth back into active neutrinos. This reconversion would have resulted in a high-energy flux which was not observed in the Kamiokande-II and IMB signal of SN~1987A. Although these re-flipped neutrinos are expected to be less abundant than the standard 10-MeV neutrinos emitted from the neutrinosphere, they possess considerably larger energies. Since the  cross section for neutrino-nucleon interaction scales approximately with the square of the energy, these neutrinos could produce observable signals in underground detectors~\cite{Akita:2022etk,Fiorillo:2022cdq,Telalovic:2024cot}.

In this paper, we return to the emission of right-handed neutrinos from the historical SN~1987A after almost 40 years. While the cooling bounds have been revisited several times over the last few decades with a varying degree of detail~\cite{Barbieri:1988nh,Notzold:1988kz,Ayala:1998qz,Ayala:1999xn,Kuznetsov:2007mp,Lychkovskiy:2009pm,Kuznetsov:2009zm}, the event-counting criterion has not been reassessed since the 1980s.
This prompted us to reconsider the constraints by including the production of $\nu_R$ from the scattering of $\nu_L$ on both protons and electrons, progress in SN simulations, and advancements in the description of the Milky Way (MW) magnetic field.
We find that one can draw robust and stringent bounds from SN~1987A, and that the electron neutrino magnetic dipole moment must be $\mu_{\nu_e} \lesssim 4.8 \times 10^{-13} \mu_B$ at 90\% C.L., superseding all existing terrestrial and astrophysical bounds. Our final results are summarized in Fig.~\ref{fig:bounds}, where we show our estimated uncertainty range and compare with existing bounds from other astrophysical, cosmological, and laboratory observables.

To substantiate our findings, we compute in Sec.~\ref{sec:emissivity} the emissivity of right-handed neutrinos produced by helicity-flipping processes due to both electrons and protons, and integrate it over Garching 1D models in Sec.~\ref{sec:num_model}, from which we draw energy-loss constraints that we validate against a simple informed one-zone model. Next, we turn to right-handed neutrinos reconversion in astrophysical B-fields in Sec.~\ref{sec:conversion}, and describe in detail the event spectrum expected in the IMB detector in Sec.~\ref{sec:detection}. Finally, we compute our revisited bounds from the event-counting criterion in Sec.~\ref{sec:analysis}. In Sec.~\ref{sec:existingbounds} we compare our findings with existing bounds on neutrino MDM, and we conclude with a summary in Sec.~\ref{sec:conclusions}.

\section{Sterile neutrino production in SNe\label{sec:emissivity}}

Right-handed neutrinos can be produced through the scattering of active left-handed neutrinos $\nu_L$ in an external field (see, e.g., Ref.~\cite{Raffelt:1996wa}). As we are concerned with ultra-relativistic neutrinos, both electric and magnetic fields in the laboratory frame contribute to the magnetic field in their rest frame.
The transition is mediated by the magnetic dipole interaction
\begin{equation}
  \mathcal L=\frac12\,\mu_\nu\bar\nu_R\,\sigma_{\alpha\beta}\,\nu_LF^{\alpha\beta}
 + \rm h.c.      \label{eq:lagrangian}
\end{equation}
where $\sigma_{\alpha\beta}=\frac{i}{2}\left[\gamma_\alpha,\gamma_\beta\right]$ is the antisymmetric combination of $\gamma$ matrices, $F^{\alpha \beta}$ is the strength of the electromagnetic field and $\mu_\nu$ is the magnetic moment of a given neutrino flavor; the latter is typically expressed in units of the Bohr magneton, $\mu_B=e/(2m_e)$, as $\mu_\nu=\kappa\,\mu_B$ with $\kappa$ a dimensionless number.  As we are mostly interested in the constraints ensuing from the back-conversion of $\nu_R$ in the Galactic magnetic fields, we will consider only the diagonal terms in the dipole moment matrix. The electric dipole moment $\epsilon_\nu$, defined by the interaction term $(1/2) \epsilon_{\nu}   \bar{\nu}   \sigma_{\alpha \beta} \gamma^5  \nu   F^{\alpha \beta}$, can be constrained in the same way.

In the hot and dense core of the SN, electromagnetic field fluctuations are sourced by all charged particles.  The relevant degrees of freedom are not only on-shell photons, but also off-shell excitations of the plasma that left-handed neutrinos can scatter with. The $\nu_R$ emission rate can thus be expressed in terms of a non-time-ordered correlator of the electromagnetic field,  playing the role of the occupation number of virtual photons. 
In thermodynamic equilibrium, this correlator is related by the fluctuation-dissipation theorem to the imaginary part of the retarded photon propagator, which in turn is determined by the polarization tensor of the medium. In linear response, the latter is defined by the current induced in the plasma by an electromagnetic field (see e.g. Ref.~\cite{kirzhnits1987general}), 
\begin{equation}
 J^\mu_{\rm med}(Q)=\,\Pi^{\mu}{}_{\nu}(Q)\,A^\nu(Q),
\end{equation}
where $Q^\mu=(\omega,\mathbf{\mathbf{q}})$ is the four-momentum carried by the virtual photon. In a plasma, the polarization tensor assumes the general form (see also Refs.~\cite{Altherr:1992mf,Braaten:1993jw,Raffelt:1996wa})
\begin{equation}
\Pi^{\mu}{}_{\nu}(Q)= 4 e^2\int\frac{d^3\mathbf{p}}{(2\pi)^3\,2E_\mathbf{p}}\,f_\mathbf{p}\sum_{\zeta=\pm}
\frac{2P^\mu P_\nu+\zeta\big(P^\mu Q_\nu+P_\nu Q^\mu-\delta^\mu_\nu\,P\!\cdot\!Q\big)}{\big(E_\mathbf{p}+\zeta(\omega+i0)\big)^2-E^2_{\mathbf p+\zeta\mathbf q}},
\label{eq:Pi_final2}
\end{equation}
where $f_\mathbf{p}=n_\mathbf{p}+\bar{n}_\mathbf{p}$ is the sum of particle and antiparticle distribution functions; for example, the electron contribution is from $f_\mathbf{p}=\left[\exp{\left(\frac{E_{\mathbf{p}}-\hat{\mu}_e}{T}\right)}+1\right]^{-1}+\left[\exp{\left(\frac{E_{\mathbf{p}}+\hat{\mu}_e}{T}\right)}+1\right]^{-1}$, where $\hat{\mu}_e$ is the electron chemical potential and $T=1/\beta$ the temperature. 
The entire problem reduces to determining what current do the different charged particles in the plasma produce in response to an external field; as muons constitute a negligible contribution to the charge density in the PNS~\cite{Bollig:2017lki} (see also Fig.~1 in Ref.~\cite{Caputo:2021rux}), we include  only the contributions from non-relativistic protons and relativistic electrons. In Appendix~\ref{app:polarization} we detail the derivation of the longitudinal and transverse components of the polarization tensor, which in the Coulomb gauge read $\Pi ^{00}=\Pi_L$ and $\Pi^a_a=\Pi_T$.

Without any further ado, the rate of $\nu_R$ production with momentum $\mathbf{k}$ per unit volume and time can be written in terms of $\Pi_L$ and $\Pi_T$ as
\begin{equation}
\begin{split}
\dot{\mathcal{N}}_{\nu_R,\mathbf{k}}
&= \frac{2\mu_\nu^2}{E_\mathbf{k}}
\int\!\frac{d^3\mathbf{p}}{(2\pi)^3\,2E_{\mathbf{p}}}\,
\frac{n^{\nu_L}_{\mathbf{p}}\,(P\!\cdot\! K)}{e^{\beta(k-p)}-1} \ \times \\
& \times \Bigg[
-\frac{(p+k)^2\,\mathrm{Im}\,\Pi_L(K-P)}
{\left| |\mathbf{k-p}|^2-\Pi_L(K-P)\right|^2}  +
\frac{4|\mathbf{k}\times\mathbf{p}|^2\,\mathrm{Im}\,\Pi_T(K-P)}
{|\mathbf{k-p}|^2\left|(K-P)^2+\Pi_T(K-P)\right|^2}
\Bigg].
\end{split}
\label{eq:rate}
\end{equation}

Leveraging the expressions for the longitudinal and transverse components of $\Pi^{\mu}{}_{\nu}$, we obtain in Appendix~\ref{app:emissivity} explicit expressions for the contributions from electrons and protons. Denoting by $\omega$ the energy exchanged with the plasma and by $E_{\nu}=|\mathbf{k}|$ the momentum of the emitted right-handed neutrino, we find the contribution to the emissivity due to virtual fluctuations sourced by relativistic electrons to be
\begin{equation}
\begin{split}
\dot{\mathcal{N}}_{\nu_R,\mathbf{k}}^{\rm electrons} =\frac{3\mu_\nu^2\omega_{\rm pl,e}^2}{4\pi}
\bigg\{ & \int_0^{+\infty} d\omega \, n^{\nu_L}_{E_{\nu}+\omega} \ +  n^{\nu_L}_{E_{\nu}} \, T \log\!\left( \frac{T E_{\nu}}{\omega_{\mathrm{pl},e}^2} \right) \bigg\},
\end{split}
\label{eq:electron_emissivity}
\end{equation}
where $\mu_\nu$ is the
neutrino magnetic moment,
and the function $n^{\nu_L}$ is the Fermi-Dirac distribution of the left-handed neutrinos; the electron plasma frequency 
\begin{equation}
    \omega_{\mathrm{pl},e}^2 = \frac{4\alpha}{3\pi}\left(\hat{\mu}_e^2 +\frac{1}{3}\pi^2T^2\right)
    \end{equation}
regulates the infrared divergence of the logarithm.

As for protons, the calculation turns out to be much simpler. They are non-relativistic and their recoil is
negligible. We can neglect the contribution from transverse virtual photons from the outset and we can assume throughout that $|\omega| \ll q$, which corresponds to the small-recoil approximation for collisions with heavy nucleons. The resulting right-handed neutrino emissivity due to proton scattering then reads
\begin{equation}
\dot{\mathcal{N}}_{\nu_R,\mathbf{k}}^{\rm protons}=\frac{\mu_\nu^2\,T\,n^{\nu_L}_{\mathbf k}\,k_{D,p}^2}{4\pi}\,\log\!\left(1+\frac{4k^2}{k_{D,p}^2}\right).
\label{eq:proton_emissivity}
\end{equation}
Here, $k_{D,p}^2$ denotes the proton contribution to the Debye screening scale, which corresponds to the static response of the medium $k_{D,p}^2=-\Pi_L(0,q)$ \cite{Raffelt:1987np}.
The same result can be recovered semiclassically  by treating the protons as static classical charges and the neutrino as a quantum particle scattering elastically off their electrostatic field. The Debye scale---that we compute neglecting the proton mild degeneracy at the center---thus acts as an infrared regulator on the momentum transfer, as in Primakoff emission \cite{Raffelt:1987np}.

Our total emissivity is given by the sum over these two contributions,
\begin{equation}
    \dot{\mathcal{N}}_{\nu_R,\mathbf{k}}=    \dot{\mathcal{N}}_{\nu_R,\mathbf{k}}^{\rm electrons}+\dot{\mathcal{N}}_{\nu_R,\mathbf{k}}^{\rm protons}.
\end{equation}
By inspecting the two contributions to the total emissivity from electrons and protons,
we find that the electron contribution Eq.~\eqref{eq:electron_emissivity} is roughly comparable to Eq.~\eqref{eq:proton_emissivity} for the conditions of the PNS. This can be contrasted with Refs.~\cite{Notzold:1988kz} and~\cite{Barbieri:1988nh}, where respectively only protons and electrons were included (in kinetic theory approximation); 
and Refs.~\cite{Ayala:1998qz,Ayala:1999xn}, that accounted only for the contributions from electrons in the context of imaginary-time and real-time thermal field theory frameworks. Therefore, our analytical results agree with the numerical findings of Refs.~\cite{Kuznetsov:2007mp,Kuznetsov:2009zm}, in which the production rates from both electrons and protons were first compared.

\section{Numerical SN models and cooling bound\label{sec:num_model}}

SN~1987A was a Type II SN at the edge of the Tarantula Nebula. It followed the collapse of the blue supergiant Sanduleak $-69^\circ$ 202a, hosted by the Large Magellanic Cloud (LMC) which in turn is a dwarf satellite galaxy of the MW. For our numerical analysis we describe SN~1987A employing the Garching 1D simulations \textsc{SFHo-18.8} and \textsc{LS220-20.0}, corresponding to a ``cold'' and a ``hot'' model respectively \cite{CCSNArchive}, both evolved with the \textsc{Prometheus-Vertex} code. These cases refer to two SN simulations, which differ in equation of state (EoS) of nuclear matter and in the mass of the compact remnant, bracketing the range of conditions that might have been realized in the core of SN~1987A, with internal temperatures reaching about $40$ and $60\rm \, MeV$, for the cold and hot model respectively \cite{Oertel:2016bki,Bollig:2020xdr, Fiorillo:2022cdq}. In both cases the electron-neutrino chemical potential at the center is $\hat{\mu}_{\nu_e}\sim 150 \,\rm  {MeV}$ and the lepton-rich core extends over a radius of $10$--$20\,\rm {km}$. The cold model releases a total energy 
$E_{\mathrm{tot,cold} } = 1.98 \times 10^{53}\,\mathrm{erg}$, while for the hot model, $E_{\mathrm{tot,hot} } = 3.93 \times 10^{53}\,\mathrm{erg}$~\cite{Fiorillo:2022cdq}. 

We begin by reassessing the SN~1987A cooling bounds~\cite{Raffelt:1996wa,Raffelt:2006cw}. The observation of the SN~1987A neutrino signal at Kamiokande-II~\cite{Kamiokande-II:1987idp,Hirata:1988ad,Hirata:1991td,Koshiba:1992yb}, IMB~\cite{Bionta:1987qt,IMB:1988suc}, and BUST~\cite{Alekseev:1987ej,Alekseev:1988gp}
broadly confirmed the theoretically expected flux~\cite{Bethe:1985sox,Janka:2012wk,Burrows:2020qrp}. The duration of several seconds and the observed energy are incompatible with excessive energy loss in hypothetical new forms of radiation such as e.g. QCD axions~\cite{Mayle:1987as,Burrows:1988ah,Raffelt:1990yz}. In principle, to obtain such bounds, one should implement the additional cooling channel in a numerical simulation of SN~1987A for a plausible range of input assumptions concerning the PNS, derive the expected neutrino signal, and compare it with the data. In practice, it is customary to obtain a rough estimate of the excluded parameter space by requiring the luminosity in novel feebly interacting particle (FIP) to be smaller than the neutrino luminosity evaluated at $\tau=1\,\rm s$ after core-bounce, which in our case would be $L_{\nu_R}(1\,{\rm s})\lesssim L_{\nu_L}(1\,\rm s)$~\cite{Raffelt:2006cw,Caputo:2021rux}.
\begin{figure}
    \centering
    \includegraphics[width=1\linewidth]{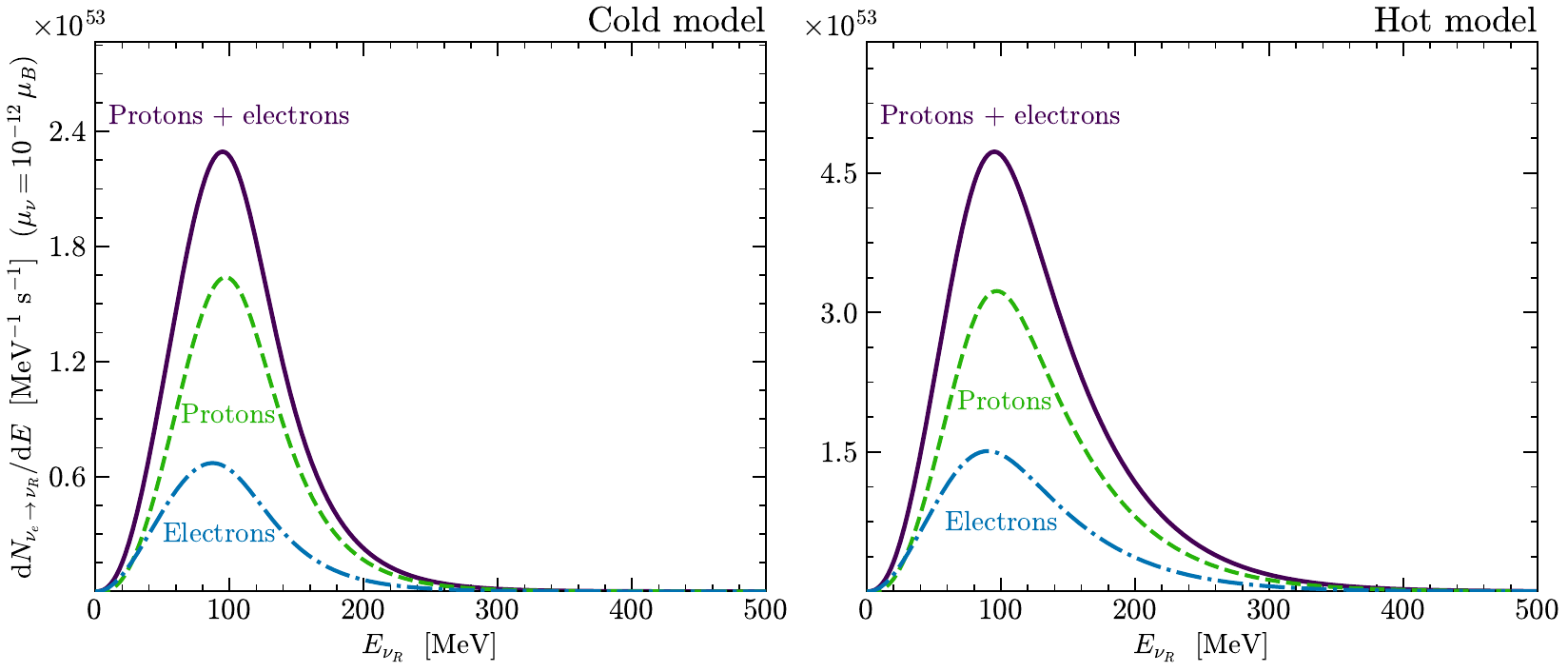}
   \caption{Energy spectrum of right-handed neutrinos for a distant observer produced through the dipole moment conversion $\nu_e \to \nu_R$ at $\tau = 1$~s after the bounce, assuming
$\mu_\nu = 10^{-12}\,\mu_B$. The \textit{left} and
\textit{right} panels show the spectra for the cold and hot models,
respectively.  The dashed green and dot-dashed blue lines show the proton
and electron contributions to the emissivity, respectively; the solid
purple line is their sum.}
    \label{fig:cold_hot_spectra_1s}
\end{figure}

We integrate the emissivities in Eqs.~\eqref{eq:electron_emissivity} and~\eqref{eq:proton_emissivity}. When computing the contributions from local emission to the luminosity arriving at large distances, one needs to include the gravitational redshift, defined by the redshifted energy $E_\nu=|\mathbf{k}|/(1+z)$ and tabulated in the models; including the total factor $(1+z)^{-2}$~\cite{Caputo:2021rux,Caputo:2022mah} from the energy redshift and time dilation, the luminosity reads
\begin{equation}
    L_{\nu_R}=\int dV \frac{1}{(1+z)^2}\int\frac{d^3\mathbf{k}}{(2\pi)^3}    \dot{\mathcal{N}}_{\nu_R,\mathbf{k}}|\mathbf{k}|\,.
\end{equation}
We show the neutrino energy spectra for the two models in Fig.~\ref{fig:cold_hot_spectra_1s}. For the cold model, summing over neutrinos and antineutrinos, the luminosity of right-handed neutrinos at $\tau=1\,\mathrm{s}$ is $L_{\nu_R} = 4.01\times10^{51}\,(\mu_\nu/10^{-12}\mu_B)^2\,\mathrm{erg\,s^{-1}}$,  which we compare with $
L_\nu(1\,\mathrm{s}) = 4.40 \times 10^{52}\,\mathrm{erg/s}$~\cite{Fiorillo:2022cdq}, leading to the cooling bound
\begin{equation}
    \mu_{\nu_e} \lesssim 3.3 \times 10^{-12}\ \mu_B.
    \label{coolingbound_cold}
\end{equation}
For the hot model, the luminosity is larger, owing to the larger temperatures reached. We find $L_{\nu_R} = 1.46\times10^{52}\,(\mu_\nu/10^{-12}\mu_B)^2\,\mathrm{erg\,s^{-1}}$, to be compared with
$ L_{\nu}(1s) = 8.29 \times
10^{52} \ \rm  erg/s  \label{bound su mu hot numerico}$~\cite{Fiorillo:2022cdq}.
Therefore, in this case, the bound reads
\begin{equation}
\mu_{\nu_e} \lesssim 2.4\times 10^{-12}\  \mu_B.
 \label{coolingbound_hot}
\end{equation}

Although they are mostly independent from the SN model, the right-handed neutrino cooling bounds can suffer from several large uncertainties, some of which are shared also by QCD axion bounds. The cooling constraints have indeed been questioned~\cite{Bar:2019ifz} as they would be hindered if SN~1987A were after all not a CCSN and no PNS was formed in its collapse. While the collapse-induced thermonuclear
explosion mechanism discussed in Ref.~\cite{Bar:2019ifz} is challenged both by observations of the SN~1987A remnant~\cite{Cigan:2019shp,Page:2020gsx,Fransson:2024csf}, and the results are possibly inconclusive about the true fate of the neutrino signal duration (as no simulation beyond $\tau=0.2\,\rm s$ was realized, see also the criticisms in Ref.~\cite{Caputo:2024oqc}), the inclusion of the additional cooling channel might still have a non-perturbative effect on the PNS larger than expected, as it could affect convection---recently found to be more relevant than initially thought (see e.g. Ref.~\cite{Fiorillo:2023frv}). Another issue with applying the QCD axion argument verbatim to right-handed neutrinos is that, like in the majoron case~\cite{Fiorillo:2022cdq}, particle emission is highest at a time preceding $\tau=1\,\rm s$. The event-counting criterion is instead robust against these systematics, so we can neglect them in the following.

\begin{figure}
    \centering
    \includegraphics[width=1\linewidth]{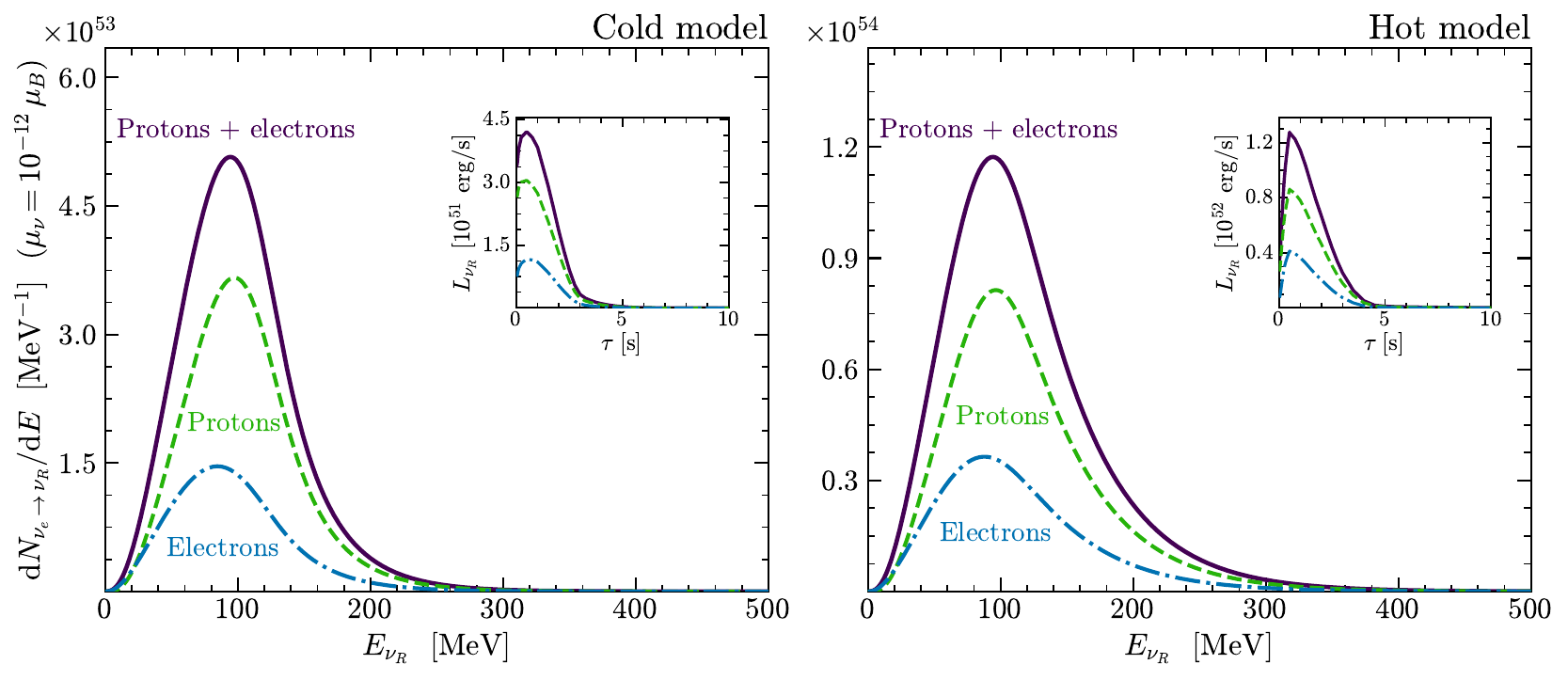}
   \caption{Time-integrated energy spectrum of right-handed neutrinos for a distant observer from
$\nu_e \to \nu_R$ conversions, for the cold
(\textit{left}, SFHo-18.8) and hot (\textit{right}, LS220-20.0) supernova models, assuming
$\mu_\nu = 10^{-12}\,\mu_B$. The dashed green and dot-dashed blue lines show the proton
and electron contributions to the emissivity, respectively; the solid
purple line is their sum.
The insets show the time evolution of the
corresponding $\nu_R$ luminosity $L_{\nu_R}$.}
    \label{fig:cold_hot_spectra}
\end{figure}

\begin{figure}
\centering
\includegraphics[width=\textwidth]{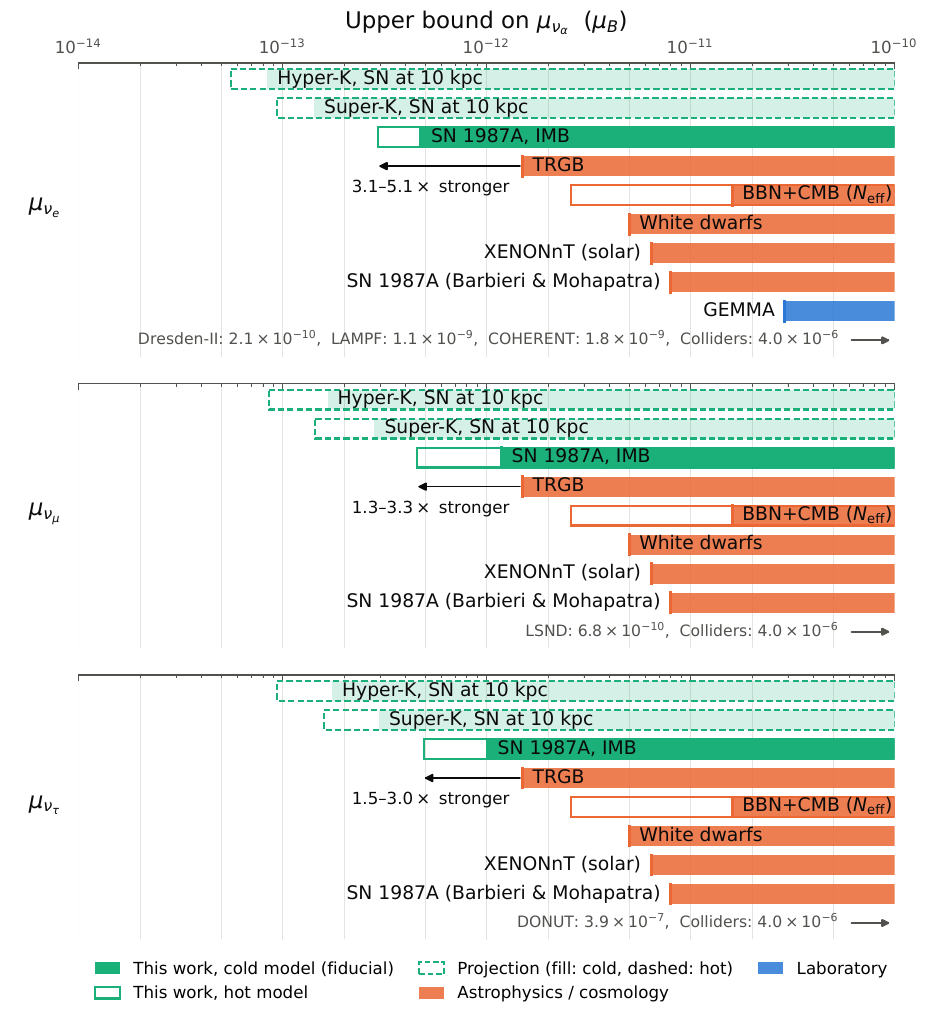}
\caption{Upper bounds on the diagonal magnetic dipole moments $\mu_{\nu_e}$, $\mu_{\nu_\mu}$ and $\mu_{\nu_\tau}$. Details and references for all bounds are given in Table~\ref{tab:macroclass_limits}.}
\label{fig:bounds}
\end{figure}

Integrating the emissivity over the volume and the entire cooling period, we obtain the differential spectrum of $\nu_R$,
\begin{equation}
    \frac{d N_{\nu_R}}{dE_{\nu}}=
    \int dt \int\,dV \frac{(1+z)^2E_{\nu}^2}{2\pi^2}
    \dot{\mathcal{N}}_{\nu_R,\mathbf{k}}\left|_{|\mathbf{k}|=(1+z)E_\nu}
    \right.,
\end{equation}
shown in Fig.~\ref{fig:cold_hot_spectra}. The right-handed neutrinos could eventually convert back into left-handed states before reaching Earth: the very same dipole moment that triggers their production in the PNS also allows for the reverse process in the presence of the Galactic magnetic field, making them detectable at terrestrial laboratories.

\section{Sterile-to-active neutrino conversion in the Galactic magnetic field\label{sec:conversion}}

The right-handed neutrinos and antineutrinos produced by helicity flip in the PNS are sterile and leave the SN freely. On their way to Earth, the component of the Galactic magnetic field orthogonal to their momentum makes their spin precess, flipping a fraction of them back to the active, left-handed helicity state~\cite{Barbieri:1988nh,Notzold:1988kz,Lattimer:1988mf,Raffelt:1996wa,Kopp:2022cug}. For ultrarelativistic neutrinos, such as those produced in SNe, helicity and chirality coincide (up to factors of order $m_\nu/E_\nu\ll1$, the ratio between the mass and the energy of the neutrino), so we use the two terms interchangeably. In this section we compute the fluence of reconverted neutrinos arriving at IMB, which we will later use to compute the expected number of events in the detector and to set upper bounds on the MDMs.

We consider Dirac neutrinos with magnetic moment matrix $\mu_\nu$ in the mass basis; the terms $\mu_{\nu_i}\equiv \mu_{ii}$ are neutrino dipole moments, while the off-diagonal terms are transition moments. Along the trajectory, parametrized by the coordinate $s$, the evolution of the system in the basis $(\nu_i^L,\nu_i^R)$, $i=1,2,3$ and in the lab frame is governed by the Hamiltonian~\cite{Fujikawa:1980yx,Raffelt:1996wa}
\begin{equation}
    H(s)=\begin{pmatrix} \dfrac{M^2}{2E_\nu}+U^\dagger V U & \mu_\nu B^*(s)\\[4pt] \mu_\nu B(s) & \dfrac{M^2}{2E_\nu}\end{pmatrix},
    \label{eq:H_full}
\end{equation}
where $M^2={\rm diag}(m_1^2,m_2^2,m_3^2)$ is the mass matrix, $V$ is the matter potential and $B(s)=B_1(s)+iB_2(s)$ combines the two components of the Galactic magnetic field perpendicular to the line of sight, $B_1$ and $B_2$; the longitudinal component does not cause helicity flips. The potential $V$ is diagonal in the flavor basis,
\begin{equation}
    V=\begin{pmatrix} V_{\rm CC}+V_{\rm NC} &0&0\\
    0& V_{\rm NC}&0 \\
    0& 0& V_{\rm NC}
    \end{pmatrix},
    \label{eq:H_full2}
\end{equation}
with $V_{\rm CC}=\sqrt{2}G_{\rm F}\mathcal{N}_e$, $V_{\rm NC}=-G_{\rm F}\mathcal{N}_n/\sqrt{2}$; here, $G_{\rm F}=1.166 \times 10^{-5} \,\rm GeV^{-2}$ is the Fermi coupling constant~\cite{Mohr:2024kco}, $\mathcal{N}_e$ and $\mathcal{N}_n$ are the free-electron and neutron number densities, and the potential acts exclusively on the active left-handed states. 

For helicity flips in the interstellar medium (ISM), a few approximations simplify the problem. First, the matter potential is negligible: for $\mathcal{N}_e\sim1.2\,{\rm cm^{-3}}$ and $\mathcal{N}_n\sim0.2\,{\rm cm^{-3}}$, obtained from the average hydrogen density $n_{\rm H}\sim1\,{\rm cm^{-3}}$ in the Galactic disk and helium density $n_{\rm He}\simeq0.1\,n_{\rm H}$~\cite{Draine:2011}, $|V|\lesssim10^{-37}\,{\rm eV}$, many orders of magnitude smaller than the mixing term $\mu |B|\simeq6\times10^{-28}\,{\rm eV}\,(\mu/10^{-13}\mu_B)(|B|/\mu{\rm G})$. Second, the vacuum term $m_i^2/2E$ is the same for $\nu_i^L$ and $\nu_i^R$, so it does not split the two helicity states of the same mass eigenstate. It does, however, split different mass eigenstates: since $\Delta m^2_{ij}/2E_\nu\gtrsim4\times10^{-13}\,{\rm eV}\gg\mu_{ij}|B|$ for $E_\nu\sim100\,{\rm MeV}$, transitions $\nu_i^L\to\nu_j^R$ with $i\neq j$ are suppressed by $(2E_\nu\mu_{ij}|B|/\Delta m^2_{ij})^2\sim10^{-30}$. For these reasons, each mass eigenstate behaves like an independent, degenerate two-level system, which can change state only due to the diagonal MDM $\mu_{ii}$.
With these assumptions, each mass eigenstate evolves according to the Schr\"odinger equation
\begin{equation}
    i\frac{d}{ds}\begin{pmatrix}\nu_i^L\\ \nu_i^R\end{pmatrix}
    =\begin{pmatrix} 0 & \mu_{ii}\,B^*(s)\\ \mu_{ii}\,B(s) & 0\end{pmatrix}
    \begin{pmatrix}\nu_i^L\\ \nu_i^R\end{pmatrix},
    \label{eq:evolution}
\end{equation}
where we dropped the term $m_i^2/2E$, common to both helicity states, which only contributes to a global phase. Therefore, this is formally equivalent to the axion-photon conversion~\cite{Brockway:1996yr,Grifols:1996id,Payez:2014xsa,Hoof:2022xbe,Fiorillo:2025gnd,Candon:2025sdm,Fiorillo:2026rgo}.

The physical parameters that govern the helicity reconversion in a magnetic field are the 3 diagonal MDMs in the mass basis, $\mu_{ii}$, while the helicity flip in the PNS involves flavor states. The two bases are related by $\mu_{\alpha\beta}=U_{\alpha i}\,\mu_{ij}\,U^*_{\beta j}$, i.e.\ $\mu_{\rm mass}=U^\dagger\mu_{\rm flavor}U$, where $U$ is the PMNS matrix. We parametrize it with the best-fit parameters of NuFit-6.0~\cite{Esteban:2024eli} for normal ordering, without SK data. For Dirac neutrinos, particle and antiparticle moments are equal in magnitude, $|\bar\mu_{ii}|=|\mu_{ii}|$ and $|\bar\mu_{\alpha\beta}|=|\mu_{\alpha\beta}|$.
We turn on one diagonal moment $\mu_{\nu_\alpha}$ at a time, in which case $\mu_{ii}=|U_{\alpha i}|^2\mu_{\nu_\alpha}$. A non-zero $\mu_{\nu_\alpha}$ therefore allows both $\nu_\alpha$ and $\bar\nu_\alpha$ to leave the PNS as right-handed and to be partially reconverted to left-handed in the Galactic magnetic field. 

To first order in perturbation theory, the probability for a mass eigenstate $i$ to flip its helicity over a distance $D$ due to Eq.~\eqref{eq:evolution} is
\begin{equation}
    P_i=|\mu_{ii}|^2 \,\left|\int_D ds\,B(s)\,\right|^2.
    \label{eq:Probability}
\end{equation}
We adopt the Galactic magnetic field model of Unger and Farrar~\cite{Unger:2023lob}, in particular the ``$\texttt{base}$" model, and include only its regular component. The values of $B_1(s)$ and $B_2(s)$ are obtained by projecting the components of the Galactic magnetic field on the plane orthogonal to the line of sight to SN~1987A, and they are the same as in Fig. 6 of \cite{Fiorillo:2025gnd}. Integrating this magnetic field over the luminosity distance $D = d_{\rm SN} = 49.6\,\rm kpc$~\cite{Pietrzynski:2019cuz}, Eq.~\eqref{eq:Probability} gives 
\begin{equation}
    P_i=7.2\times10^{-2}\,\left|\frac{\mu_{ii}}{10^{-13}\,\mu_B}\right|^2 .
    \label{eq:computed_P}
\end{equation}
This estimate, obtained perturbatively, is valid only as long as $P_i\ll1$. For the MDMs relevant in this work, this is not always the case; for instance, for $\mu_{\nu_\mu}=7\times10^{-13}\,\mu_B$, Eq.~\eqref{eq:computed_P} would give $P_3\simeq1$. 

For this reason, we solve Eq.~\eqref{eq:evolution} numerically. We divide the line of sight into short steps $\Delta s$, in which the transverse field $B(s)$ can be taken as constant. Within each step, the Hamiltonian of Eq.~\eqref{eq:evolution} is a constant $2\times2$ matrix, with eigenvalues $\pm\mu_{ii}| B|$: by diagonalizing it, we obtain the exact evolution operator over that step. The evolution operator along the whole line of sight is the ordered product of the evolution operators of all steps, from SN~1987A to Earth. Its off-diagonal element is the amplitude for a right-handed state to arrive as left-handed, so the flip probability $P_i$ is its squared modulus. For small $\mu_{ii}$ this reproduces Eq.~\eqref{eq:computed_P}. The perturbative result overestimates the signal: it would make the bounds stronger by a few percent for the hot model, and by $10-40\%$ in the cold model, depending on the bound. In the following, $P_i$ always denotes the exact flip probability.

The time-integrated fluence at Earth of right-handed neutrinos of flavor $\alpha$ emitted by the PNS is
\begin{equation}
    F^\alpha_{\nu_R}(E_{\nu})=\frac{1}{4\pi d_{\rm SN}^2}\frac{dN_{\nu_R}^\alpha}{dE_{\nu}}.
\end{equation}
Assuming that neutrinos free-stream out of the PNS once they convert to right-handed (see Sec.~\ref{sec:emissivity}), $F^\alpha_{\nu_R}\propto\mu_{\nu_\alpha}^2$. Over the distance $D$, the oscillation phases $\Delta m^2_{ij}D/2E_\nu$ average out, so the mass eigenstates arrive as an incoherent superposition. Projecting onto the mass basis and applying the flip probability $P_i$, evaluated at $\mu_{ii}=|U_{\alpha i}|^2\mu_{\nu_\alpha}$, the left-handed fluence of flavor $\beta$ at the detector is
\begin{equation}
    \tilde F_{\beta}(E_\nu)=\sum_i P^\oplus_{i\to\beta}(E_\nu)\,P_i\,|U_{\alpha i}|^2\,F^{\alpha}_{\nu_R}(E_\nu),
    \label{eq:fluence_earth}
\end{equation}
where $P^\oplus_{i\to\beta}(E_\nu)$ is the probability that a mass eigenstate $\nu_i$ reaching the Earth is detected as $\nu_\beta$. In vacuum, $P^\oplus_{i\to\beta}=|U_{\beta i}|^2$. However, at the time of the burst SN~1987A was $42^\circ$ below the horizon of IMB~\cite{Fiorillo:2023frv}, so that neutrinos crossed $8600\,\rm km$ of the Earth's mantle before reaching the detector. In the mantle, the matter potential $V_{CC}=\sqrt2G_F\mathcal{N}_e\approx10^{-13}\,\rm eV$ is comparable to $\Delta m^2_{21}/2E_\nu$ for $E_\nu\sim100\,\rm MeV$, which means that the reconverted neutrinos lie in the energy range of the Mikheyev–Smirnov–Wolfenstein resonance of the solar splitting. We compute $P^\oplus_{i\to\beta}$ by solving the three-flavor evolution along the chord through the Earth, with the density profile of the Preliminary Reference Earth Model (PREM)~\cite{DZIEWONSKI1981297}. As for the helicity flip, we divide the chord into short steps of constant density, diagonalize the Hamiltonian $H=U\,{\rm diag}(0,\Delta m^2_{21},\Delta m^2_{31})\,U^\dagger/2E_\nu+{\rm diag}(V,0,0)$ in each step, and multiply the resulting evolution operators; for antineutrinos, $U\to U^*$, $V\to-V$. The $\nu_3$ component is almost unaffected, since $2E_\nu V/\Delta m^2_{31}\lesssim0.03$. Instead, above the resonance a large fraction of $\nu_1$ is detected as $\nu_\mu$ or $\nu_\tau$: $P^\oplus_{1\to e}$ drops from $|U_{e1}|^2=0.68$ to $0.2-0.4$ between $100$ and $250\,\rm MeV$. For antineutrinos the effect has the opposite sign, and $P^\oplus_{\bar1\to\bar e}$ increases up to $\simeq0.9$. Eq.~\eqref{eq:fluence_earth} holds for antineutrinos with the same flip probabilities $P_i$, but different $P^\oplus_{i\to\beta}$ and spectrum, $F_{\nu_R}^{\bar\alpha}\neq F_{\nu_R}^{\alpha}$. For each $\mu_{\nu_\alpha}$, the fluence at the detector is the sum of the reconverted neutrino and antineutrino fluences.

\subsection{Additional contributions to the conversion probability}

Along the line of sight, all magnetic fields with components perpendicular to the $\nu_R$ momentum contribute to the sterile-active conversion. Right-handed neutrinos also cross the magnetic fields of the PNS and of the progenitor star. In principle, this would induce additional helicity flips with amplitude proportional to $\sim BR$, where $R$ is the radius of the PNS. For a PNS this value could exceed the integral along the Galactic line of sight, $|\int B\,ds|\simeq10^{16}\,{\rm G\,cm}$;  assuming $R\simeq 10\,\rm km$ and $B\simeq 10^{12}-10^{15}\,\rm G$ one finds $BR\sim10^{18}-10^{21}\,\rm G\,cm$. The resulting active neutrinos would also contribute to the explosion, as proposed by Dar~\cite{Dar:1987yv}. (See  related discussions in Refs.~\cite{Nussinov:1987zr,Goldman:1987fg,Voloshin:1988xu,Okun:1988qs,blinnikov1988models,Athar:1995cx,Anikin:2010rh}.) As noted in Refs.~\cite{Notzold:1988kz,Lattimer:1988mf}, in the dense environment of the PNS this term is much smaller than the matter potential $V$ of the active states, which reads $V\approx10^{-2}\,\rm eV \left(\frac{\rho}{10^{11}\rm g\,cm^{-3}}\right)$. The helicity flip inside the PNS due to interaction with the magnetic field is therefore suppressed by a factor $(2\mu_\nu B/V)^2\lesssim10^{-7}$ at $\mu_\nu = 10^{-13} \mu_B$; while this factor can become larger as the density diminishes after the explosion, it remains negligible, barring extremely large $B$ fields associated with the formation of a magnetar. Likewise, one can neglect the contribution to conversion from the magnetic field at the progenitor surface (see a similar discussion in the context of axion-photon conversion in Ref.~\cite{Fiorillo:2025gnd}).

Besides the ``$\texttt{base}$" model, Ref.~\cite{Unger:2023lob} provides seven variants that fit the same data with different assumptions on the disk and halo fields, the thermal and cosmic-ray electron densities, and the synchrotron data; their spread provides an estimate of the systematic uncertainty of the model. Along the line of sight to SN~1987A, $|\int ds\,B(s)|^2$ ranges from $0.82$ (``\texttt{twistX}") to $1.8$ (``\texttt{nebCor}") times its value in the ``\texttt{base}" model. Since for small moments the signal grows as $\mu^4|\int ds\,B|^2$, the bounds depend only weakly on the field, as $|\int ds\,B|^{-1/2}$, and even more weakly when the conversion approaches saturation. Repeating our analysis for all eight models, the bounds change by between $-13\%$ and $+4\%$ with respect to the ``\texttt{base}" model, which we keep as fiducial.

One more field contributing to the integral along the line of sight is the LMC coherent one; as shown in
Ref.~\cite{Fiorillo:2026rgo}, such a coherent field results in a correction at the level of tens percent on the conversion probability. However, the turbulent component of the Galactic magnetic field can dominate the average conversion probability. Nevertheless, once  fluctuations in the conversion probability along a single line of sight are taken into account, the marginalization over turbulent realizations changes the bounds negligibly.

\section{Active neutrino detection in IMB\label{sec:detection}}

Having computed the fluences at Earth as functions of the three MDMs $\mu_{\nu_\alpha}$, we can compute the resulting number of expected events in the IMB detector; as the constraints are mainly determined by the fiducial volume, we discard the data of Kamiokande-II, which would strengthen the bounds at the percent level. These fluences correspond to the (anti)neutrinos that left SN~1987A as right-handed and were reconverted to active states while propagating in the Galactic magnetic field.

The IMB experiment was a water Cherenkov detector with a fiducial mass of $6.8\,\rm kton$~\cite{IMB:1988suc}. This corresponds to $N_p = 4.5 \times 10^{32}$ hydrogen nuclei and $N_O = N_p/2 $ oxygen nuclei. The detector deployed photomultiplier tubes (PMTs) to measure the Cherenkov light emitted by charged leptons produced in neutrino interactions.

At typical SN energies, the dominant detection channel is inverse beta decay (IBD) of electron antineutrinos on free protons, $\bar{\nu}_e + p \to e^+ + n$. Neglecting nucleon recoil, the positron carries an energy $E_{e^+} = E_\nu - Q_{\bar{\nu}_e p}$, with $Q_{\bar{\nu}_e p} = 1.29\,{\rm MeV}$. Above $\simeq 70\,{\rm MeV}$, charged-current (CC) interactions on oxygen become the most effective channel: $\nu_e + {}^{16}{\rm O} \to e^- + {}^{16}{\rm F}^*$ and $\bar{\nu}_e + {}^{16}{\rm O} \to e^+ + {}^{16}{\rm N}^*$. The resulting $e^\pm$ carry an energy $E_{e^\pm} = E_\nu - Q_k$, with $Q_{\nu_e{\rm O}} = 15.4\,{\rm MeV}$ and $Q_{\bar{\nu}_e{\rm O}} = 11.4\,{\rm MeV}$. For each of these channels $k$, with $N_k$ targets and cross section $\sigma_k$, the spectrum of the $e^\pm$ produced in the detector is
\begin{equation}
    \frac{dN_k}{dE_{e^\pm}} = N_k\,\tilde F_{k}(E_{e^\pm}+Q_k)\,\sigma_k(E_{e^\pm}+Q_k),
    \label{eq:spectrum_e}
\end{equation}
where $\tilde{F}_k$ is the reconverted fluence of the flavor that induces the reaction, $\tilde{F}_e$ or $\tilde{\bar F}_{e}$.

For $E_\nu\gtrsim110\,\rm MeV$, also the CC interactions of muon (anti-)neutrinos become kinematically possible: $\nu_\mu+{}^{16}\rm O\to\mu^-+X$, $\bar{\nu}_\mu+p\to\mu^++n$ and $\bar{\nu}_\mu+{}^{16}\rm O \to \mu^+ + X$, where $X$ is the final hadronic state. The resulting $\mu^\pm$ are quickly stopped due to the high ionization losses, and within a length of $\mathcal{O}(1\,\rm m)$ they decay at rest into $e^\pm$. These electrons and positrons are produced with an energy distribution given by the Michel spectrum
\begin{equation}
    \frac{dN_{e^\pm}}{dE_{e^\pm}} = \frac{4}{m_\mu}\left(\frac{2E_{e^\pm}}{m_\mu}\right)^2 \left(3 - \frac{4E_{e^\pm}}{m_\mu}\right),
    \label{eq:Michel}
\end{equation} 
and stops at the kinematic maximum $E_{e^\pm} = m_\mu/2$.

Equation~\eqref{eq:Michel} is normalized to unity, so the energy distribution of the events induced by muon interactions have to be multiplied by the total number of muon-induced events, as
\begin{equation}
    \frac{dN^\mu}{dE_{e^\pm}} = \frac{dN_{e^\pm}}{dE_{e^\pm}} \sum_{k(\mu^\pm)}\int dE_\nu\,\tilde F_k(E_\nu)\,\sigma_k(E_\nu)
    \label{eq:muon_events}
\end{equation}

The cross sections of the processes considered are taken from Refs.~\cite{Marteau:1999zp,Kolbe:2002gk,Strumia:2003zx,Formaggio:2012cpf}. The electrons and positrons produced are then detected with the energy-dependent efficiency $\epsilon(E_{e^\pm})$ shown in the Supplemental Material of Ref.~\cite{Fiorillo:2022cdq} (see also Ref.~\cite{Jegerlehner:1996kx}).

Since IMB could not distinguish the different detection channels, we sum over the six interaction channels considered here, weighting each by $\epsilon(E_{e^\pm})$, to obtain the total differential distribution of electrons and positrons as they are detected in IMB,
\begin{equation}
    \frac{dN}{dE_{e^\pm}} = \epsilon(E_{e^\pm})\left[\sum_{k(e)} \frac{dN_k}{dE_{e^\pm}}+\frac{dN^\mu}{dE_{e^\pm}}\right],
\end{equation}
where the sum is over the reactions $k(e)$ involving electron (anti)neutrinos, and the contribution from muon (anti-neutrinos) is given by Eq.~\eqref{eq:muon_events}.

An $e^\pm$ will be observed in the detector with a reconstructed energy $E_{\rm det}$, which depends on the number of PMTs triggered by its Cherenkov light. We model the probability for an $e^\pm$ of true energy $E_{e^\pm}$ to be detected with energy $E_{\rm det}$ as in Ref.~\cite{Jegerlehner:1996kx}, assuming the Gaussian distribution
\begin{equation}
    P(E_{\rm det}|E_{e^\pm}) = \frac{1}{\sqrt{2\pi}\sigma(E_{e^\pm})}\exp\left(-\frac{(E_{\rm det}-E_{e^\pm})^2}{2\sigma^2(E_{e^\pm})}\right),
\end{equation}
where $\sigma(E_{e^\pm})$ is an energy-dependent dispersion estimated as $\sigma(E_{e^\pm}) = \sqrt{E_{e^\pm}E_\sigma}$, with $E_\sigma = 1.35\, \rm MeV$ being the energy resolution of the IMB experiment.

The spectrum in reconstructed energy is then
\begin{equation}
    \frac{dN}{dE_{\rm det}}=\int dE_{e^\pm}\,P(E_{\rm det}|E_{e^\pm})\,\frac{dN}{dE_{e^\pm}}.
    \label{eq:Edet}
\end{equation}
 
\section{Analysis and results\label{sec:analysis}}
IMB did not detect any events with $E_{\rm det}\geq 40\,\rm MeV$ at the time of the neutrino arrival. Reconverted right-handed neutrinos have energies of order $100\,{\rm MeV}$, and would produce $e^\pm$ events mostly above this threshold. Their non-observation therefore allows us to set upper limits on the MDMs $\mu_{\nu_\alpha}$. For a given value of $\mu_{\nu_\alpha}$, we compute the right-handed fluences $F_{\nu_R}^\alpha$ and $F_{\nu_R}^{\bar{\alpha}}$ integrated over the full $10\,{\rm s}$ of the simulation, and propagate them to the detector using the exact flip probabilities and the flavor conversion in the Earth of Sec.~\ref{sec:conversion}. We then compute the event spectrum $dN/dE_{\rm det}$, including both neutrinos and antineutrinos, and integrate it over $E_{\rm det}\geq40\,{\rm MeV}$ to obtain the expected number of events $N(\mu_{\nu_\alpha})$. With zero observed events, and neglecting the contribution of standard SN neutrinos above $40\,{\rm MeV}$, which makes the bounds conservative, we exclude at 90\% C.L.\ the values of $\mu_{\nu_\alpha}$ for which $N(\mu_{\nu_\alpha})>2.30$. Since the exact $P_i$ saturates and oscillates, $N$ is not a power law of $\mu_{\nu_\alpha}$, although for small values of $\mu_{\nu_\alpha}$, $N\propto \mu_{\nu_\alpha}^4$, recovering the perturbative result.

The resulting upper limits are listed in Table~\ref{tab:bounds}. We take the cold model as fiducial, since it gives the most conservative bounds. The bound on $\mu_{\nu_e}$ is the strongest, since electron (anti)neutrinos reconvert mostly into electron flavor and are detected through the efficient CC channels on protons and oxygen, while for $\mu_{\nu_\mu}$ and $\mu_{\nu_\tau}$ a significant part of the signal comes from Michel electrons, which lie below $m_\mu/2=53\,{\rm MeV}$. The bounds are a factor $1.6-2.6$ stronger for the hot model, which emits more right-handed neutrinos at high energies and would require smaller values of $\mu_{\nu_\alpha}$ to reach the threshold of $2.3$ events. 

\begin{table}[htbp]
    \centering
    \begin{tabular}{lcc}
        \hline
        & \textbf{Cold} & \textbf{Hot} \\
        \hline
        $\mu_{\nu_e}\ [10^{-13}\,\mu_B]$ & $4.8$ & $3.0$ \\
        $\mu_{\nu_\mu}\ [10^{-13}\,\mu_B]$ & $12$ & $4.6$ \\
        $\mu_{\nu_\tau}\ [10^{-13}\,\mu_B]$ & $10$ & $4.9$ \\
        \hline
    \end{tabular}
     \caption{Upper bounds at 90\% C.L.\ on the diagonal Dirac MDM, defined in the flavor basis as in Sec.~\ref{sec:conversion}, for the cold and hot SN models. }
         \label{tab:bounds}
\end{table}

\subsection{One-zone estimate}
To cross-check the scaling of the constraints, we can obtain a useful analytical estimate of the numerical bounds by using a one-zone approximation of the SN core. To this end, we compute the energy-loss rate per unit volume from the
emission rates of Eqs.~\eqref{eq:electron_emissivity}
and~\eqref{eq:proton_emissivity},
\begin{equation}
 \mathcal{Q}_{\nu_R}=\int\frac{d^3\mathbf{k}}{(2\pi)^3}\,
  \dot{\mathcal{N}}_{\nu_R,\mathbf{k}}\,|\mathbf{k}|.
\end{equation}
We perform the integral analytically for both protons and electrons.
For electrons we find 
\begin{equation}
 \mathcal{Q}_e=\frac{3\,\mu_\nu^2\,\omega_{\rm pl, e}^2\,\hat{\mu}_{\nu_e}^4}{32\pi^3}
  \left[\frac{\hat{\mu}_{\nu_e}}{5}
  +T\left(\ln\frac{T\,\hat{\mu}_{\nu_e}}{\omega_{{\rm pl},e}^2}-\frac14\right)\right],
  \label{eq:Qe_onezone}
\end{equation}
where we used the degeneracy of electron neutrinos.
The proton contribution reads
\begin{equation}
 \mathcal{Q}_p=\mu_\nu^2\,\frac{T\,k_{D,p}^2\,\hat{\mu}_{\nu_e}^4}{32\pi^3}
  \left[\left(1-\frac{1}{x^2}\right)\ln(1+x)+\frac{1}{x}-\frac12\right],
  \label{eq:Qp_onezone}
\end{equation}
where $x\equiv4\hat{\mu}_{\nu_e}^2/k_{D,p}^2$ and $\hat{\mu}_{\nu_e}$ is the chemical
potential of electron neutrinos.

To evaluate the bound within the one-zone approximation, we take the
values of $T$, $k_D$, $\hat{\mu}_{\nu_e}$ and $\hat{\mu}_e$ as volume averages roughly calibrated over the SN core at
$\tau=1~\mathrm{s}$ in the cold model: $T\simeq30~\mathrm{MeV}$, $k_{D,p}\simeq30~\mathrm{MeV}$,
$\hat{\mu}_{\nu_e}\simeq100~\mathrm{MeV}$ and $\hat{\mu}_e\simeq190~\mathrm{MeV}$,
corresponding to $\omega_{{\rm pl},e}\simeq10~\mathrm{MeV}$. 
Within the one-zone model, the emitting region is modelled as a sphere of radius $R\simeq10~\mathrm{km}$; requiring---as done in Sec.~\ref{sec:num_model}---that $L_{\nu_R}= \mathcal{Q}_{\nu_R}\,\mathcal{V}\lesssim4.4\, \times 10^{52} \rm \, erg \, s^{-1},$ with $ \mathcal{Q}_{\nu_R}= \mathcal{Q}_p+ \mathcal{Q}_e$ the total energy-loss rate and $\mathcal{V}=(4\pi/3)R^3$, we obtain
\begin{equation}
  \mu_\nu \lesssim 5.3\times10^{-12}\,\mu_B\,,
\end{equation}
 where we neglected the contribution from antineutrinos.
 
Similarly to this derivation, we can obtain a rough  estimate of the event-counting bound arising from the reconversion of $\nu_R$ into $\nu_L$. For a uniform transverse field, the probability of reconversion simplifies to
\begin{equation}
    P(\nu_R \rightarrow \nu_L) \simeq \left(\mu_\nu B\,\ell\right)^2 .
\end{equation}
Here, $\ell \simeq 10\,\mathrm{kpc}$ is the distance travelled in the Galactic magnetic field and $B \simeq 0.3\,\mu\mathrm{G}$ is the transverse component of the field~\cite{Fiorillo:2025gnd, Fiorillo:2026rgo}.
The number of reconverted neutrinos is obtained by multiplying this probability by the total number of emitted right-handed neutrinos, $N_{\nu_R}$, which is deduced by integrating the proton and electron emission rates over the neutrino phase space, $ \dot{\mathcal{N}}^{\rm\,  e,  p}_{\nu_R}=\int\frac{d^3\mathbf{k}}{(2\pi)^3}\,\dot{\mathcal{N}}^{\, \rm e, p}_{\nu_R,\mathbf{k}}$, from which we get
the two contributions
\begin{equation}
\dot{\mathcal{N}}^{\rm electrons}_{\nu_R}=\frac{\,\mu_\nu^2\,\omega_{\rm pl, e}^2\,\hat{\mu}_{\nu_e}^3}{8\pi^3}
  \left[\frac{\hat{\mu}_{\nu_e}}{4}
  +T\left(\ln\frac{T\,\hat{\mu}_{\nu_e}}{\omega_{{\rm pl},e}^2}-\frac13\right)\right]
\end{equation}
for electrons and
\begin{equation}
 \dot{\mathcal{N}}^{\rm protons}_{\nu_R}=\mu_\nu^2\,\frac{T\,k_{D,p}^2\,\hat{\mu}_{\nu_e}^3}{24\pi^3}
  \left[\ln(1+x)-\frac23+\frac{2}{x}
  -\frac{2\arctan\sqrt{x}}{x^{3/2}}\right]
\end{equation}
for protons, with $x$ defined as above. 
Thus, the total number of emitted right-handed neutrinos is $N_{\nu_R}=(\dot{\mathcal{N}}^{\rm electrons }_{\nu_R}+\dot{\mathcal{N}}^{\rm protons}_{\nu_R})\,\mathcal{V}\Delta\tau$, with $\Delta \tau\simeq2~\mathrm{s}$ the duration of the emission. Multiplying by $P$ we obtain the number of re-flipped neutrinos, $N_{e}=PN^{e}_{\nu_R}$.

The cross section $\nu_e + {}^{16}{\rm O} \to e^- + {}^{16}{\rm F}^*$ is roughly proportional to the energy squared, $\sigma \simeq \bar{\sigma} E_\nu^2$ with $\bar{\sigma} \simeq 10^{-43}\,\mathrm{cm}^2/\mathrm{MeV}^2$ and $\langle E_\nu^2\rangle\simeq\frac{3}{5}\,\hat{\mu}_{\nu_e}^2$.
 The neutrino fluence at Earth is $\tilde{F}_{e} = N_e/4\pi d_{\rm SN}^2$, where $d_{\rm SN}\simeq50~\mathrm{kpc}$ denotes the distance to SN~1987A, and the total number of events in the \(\sim 70\,\mathrm{MeV}\) energy range is $N = \sigma N_O \tilde{F}_e$, namely
\begin{equation}
  N=\frac{\bar\sigma\,\langle E_\nu^2\rangle\,N_O}{4\pi d_{\rm SN}^2}\,
  \left(\mu_\nu B\,\ell\right)^2 N^e_{\nu_R}\,.
  \label{eq:Nevents_onezone}
\end{equation}
Using the same average values adopted for the cooling bound, the absence of such high-energy events, $N \lesssim 1$, yields
\begin{equation}
  \mu_{\nu} \lesssim3.3 \times 10^{-13} \mu_B
\end{equation}
as the upper bound on the neutrino magnetic dipole moment. Gravitational redshift has an impact on the signal that can be easily estimated. The redshift is given by $z\simeq GM/R\simeq0.3$, where $G$ is Newton's constant, while $M$ and $R$ are respectively the mass and the radius of the PNS. Since the detection cross section scales as $\sigma\propto E^2$, the number of detected events  decreases by a factor $1/(1+z)^3\simeq 0.46$.

 These outcomes suggest how the discrepancy between our results and bounds obtained in previous works \cite{Barbieri:1988nh, Ayala:1998qz, Kuznetsov:2007mp} can be traced back to the parameters adopted for the SN core, which
are outdated in the light of modern simulations. In particular, those works
assumed considerably larger chemical potentials
($\hat{\mu}_{\nu_e}\simeq 160~\mathrm{MeV}$),
temperatures up to $60~\mathrm{MeV}$ and a higher density
($\rho\simeq 8\times10^{14}~\mathrm{g\,cm^{-3}}$), in contrast with the average density $\rho \simeq 3\times 10^{14}\,  \rm g \, cm^{-3}$  found in modern models. 
Beyond the choice of the SN model, a further difference concerns the treatment of the plasma components---as stressed in Sec.~\ref{sec:emissivity}, in this work both the proton and the electron contributions to the production of right-handed neutrinos are taken into account.
A further difference concerns the knowledge of the Galactic magnetic field, which determines the probability that the right-handed neutrinos escaping from the SN core are converted back into detectable left-handed states on
their way to Earth. The early estimates \cite{Barbieri:1988nh} relied on a simplified description, with a uniform field of fixed strength over a given path length, in contrast with present-day models of the Galactic magnetic field \cite{Unger:2023lob}.

\subsection{Prospects for the next Galactic Supernova}
A future Galactic SN would allow the same analysis with higher statistics. We estimate the bounds that will be obtained by this event, assuming a SN at $d=10\,\rm kpc$ observed by Super-Kamiokande, with a fiducial mass $M$ of $22.5\,\rm kt$~\cite{Super-Kamiokande:2002weg}. For small MDMs the number of events scales as $\mu_{\nu_\alpha}^4M d^{-2}$, so that the sensitivity improves by a factor $[(M_{\rm SK}/M_{\rm IMB})(d_{\rm 87A}/d)^2]^{1/4}\simeq3.0$ with respect to SN~1987A. Rescaling our perturbative bounds, which is appropriate since at these smaller moments the flip probabilities are $P_i\lesssim0.17$, we find a sensitivity $\mu_{\nu_e}\lesssim1.4\times10^{-13}\,\mu_B$, $\mu_{\nu_\mu}\lesssim2.8\times10^{-13}\,\mu_B$ and $\mu_{\nu_\tau}\lesssim3.0\times10^{-13}\,\mu_B$ for the cold model, and $9.4\times10^{-14}$, $1.5\times10^{-13}$ and $1.6\times10^{-13}\,\mu_B$ for the hot model. This estimate assumes the same detector response as for IMB and the same Galactic field integral as for SN~1987A. For Hyper-Kamiokande, with one tank of $187\,\rm kt$ fiducial mass~\cite{Hyper-Kamiokande:2018ofw}, the same rescaling gives a further improvement by $(M_{\rm HK}/M_{\rm SK})^{1/4}\simeq1.7$.

A dedicated analysis would have to account for the different efficiency and energy resolution of these detectors, and for the standard SN signal, which would be observed with much more events in Super-Kamiokande, whose thermal tail would require a higher energy cut, partly removing the Michel electrons that contribute to the signal for $\mu_{\nu_\mu}$ and $\mu_{\nu_\tau}$. Moreover, the direction of the SN would affect the Galactic field integral, which could change by a factor of a few, and the matter effects in the propagation through the Earth, particularly relevant for electron (anti)neutrinos.

\begin{table*}[t]
\footnotesize
  \begin{minipage}{0.6\linewidth}
    \renewcommand{\footnoterule}{\vspace*{-8pt}} 
  \centering
  \begin{tabular}{@{} l l c c c c @{}}
\toprule
Method & Experiment / Probe & Limit [$\mu_{\mathrm{B}}$] & CL & Year & Ref. \\
\midrule
\multicolumn{6}{c}{\textbf{Laboratory Probes}} \\
\midrule
Reactor $\bar{\nu}_e$ E$\nu$ES & GEMMA & $\mu_{\nu_e} < 2.9 \times 10^{-11}$ & 90\% & 2012 & \cite{Beda:2012zz} \\
Accelerator $\nu_e$ E$\nu$ES & LAMPF & $\mu_{\nu_e} < 1.1 \times 10^{-9}$ & 90\% & 1992 & \cite{Allen:1992qe} \\
Accelerator $\nu_\mu, \bar{\nu}_\mu$ E$\nu$ES & LSND & $\mu_{\nu_\mu} < 6.8 \times 10^{-10}$ & 90\% & 2001 & \cite{LSND:2001akn} \\
Accelerator $\nu_\tau, \bar{\nu}_\tau$ E$\nu$ES & DONUT & $\mu_{\nu_\tau} < 3.9 \times 10^{-7}$ & 90\% & 2001 & \cite{DONUT:2001zvi} \\
Accelerator $\nu_e, \nu_\mu, \bar{\nu}_\mu$ CE$\nu$NS+E$\nu$ES & COHERENT & $\mu_{\nu_e} < 1.8 \times 10^{-9}$ & 90\% & 2022 & \cite{Coloma:2022avw,AtzoriCorona:2022qrf} \\
Reactor $\bar{\nu}_e$ CE$\nu$NS+E$\nu$ES & Dresden-II & $\mu_{\nu_e} < 2.1 \times 10^{-10}$ & 90\% & 2022 & \cite{Coloma:2022avw,AtzoriCorona:2022qrf} \\
$e^+ e^- \to \nu \bar{\nu} \gamma$ (Colliders) & Combined data\footnote{ASP, MAC, CELLO, MARK-J} & $\mu_{\nu} < 4.0 \times 10^{-6}$ & 90\% & 1987 & \cite{Grotch:1987un} \\
\midrule
\multicolumn{6}{c}{\textbf{Astrophysical Probes}} \\
\midrule
SN high-energy neutrino flux& SN~1987A & $\mu_{\nu_e} < 4.8 \times 10^{-13}$ & 90\% & 2026 & This work \\
SN high-energy neutrino flux& SN~1987A & $\mu_{\nu_\mu} < 1.2 \times 10^{-12}$ & 90\% & 2026 & This work \\
SN high-energy neutrino flux& SN~1987A & $\mu_{\nu_\tau} < 1.0 \times 10^{-12}$ & 90\% & 2026 & This work \\
Solar $\nu$ E$\nu$ES (Direct Detection) & XENONnT & $\mu_{\nu} < 6.4 \times 10^{-12}$ & 90\% & 2022 & \cite{XENON:2022ltv} \\
Red-Giant Branch (TRGB) Cooling & NGC 4258 & $\mu_{\nu} < 1.5 \times 10^{-12}$ & 95\% & 2020 & \cite{Capozzi:2020cbu} \\
White Dwarf Cooling & Luminosity Function & $\mu_{\nu} < 5.0 \times 10^{-12}$ & 90\% & 2014 & \cite{MillerBertolami:2014oki} \\
CCSN neutrino signal duration & SN~1987A & $\mu_{\nu} < 0.8 \times 10^{-11}$ & $-$ & 1988 & \cite{Barbieri:1988nh} \\
\midrule
\multicolumn{6}{c}{\textbf{Cosmological Probes}} \\
\midrule
BBN and CMB& $N_{\rm eff}$ & $\mu_{\nu} < 1.6 \times 10^{-11}$\footnote{Assuming a reheating temperature $T_{\rm rh}=100\,\rm MeV$} & 95\% & 2024 & \cite{Carenza:2022ngg} \\
BBN and CMB& $N_{\rm eff}$ & $\mu_{\nu} < 2.6 \times 10^{-12}$\footnote{Assuming a reheating temperature $T_{\rm rh}=100\,\rm GeV$} & 95\% & 2024 & \cite{Li:2022dkc,Carenza:2022ngg} \\
\bottomrule
\end{tabular}
\end{minipage}
\caption{Summary of constraints on neutrino dipole moments categorized into laboratory, astrophysical, and cosmological probes.}
\label{tab:macroclass_limits}
\end{table*}

\section{Comparison with other constraints\label{sec:existingbounds}}
Our novel bounds can be compared with a plethora of laboratory, astrophysical, and cosmological bounds on neutrino dipole moments (Table~\ref{tab:macroclass_limits}). Broadly speaking, astrophysical probes (e.g. from measurements of the red giant branch tip luminosity) and cosmological observations ($N_{\rm eff}$ from BBN and CMB) achieve extremely stringent constraints, while terrestrial experiments measuring coherent elastic neutrino-nucleus scattering (CE$\nu$NS) and neutrino-electron elastic scattering (E$\nu$ES) can probe the dipole moments in a controlled laboratory environment.

On Earth, the strongest constraints on neutrino dipole moments have been obtained in laboratory short-baseline experiments through E$\nu$ES with reactor antineutrinos~\cite{Derbin:1993wy, Vidyakin:1992nf, MUNU:2005xnz,TEXONO:2006xds,Beda:2009kx,Beda:2012zz, CONUS:2022qbb,Coloma:2022avw,AtzoriCorona:2022qrf}. The other flavors are best constrained by E$\nu$ES and CE$\nu$NS, which require the use of accelerator neutrinos~\cite{Allen:1992qe, Ahrens:1990fp, LSND:2001akn, Cooper-Sarkar:1991vsl,COHERENT:2017ipa,COHERENT:2020iec, COHERENT:2021xmm,Coloma:2022avw, Abdullah:2022zue, Cadeddu:2023tkp}; additional constraints ensue from searches in $e^+e^-$ collisions~\cite{Grotch:1987un, Tanimoto:2000am}.  For muon neutrinos, accelerator-based measurements from LSND \cite{LSND:2001akn} give $\mu_{\nu_\mu} < 6.8 \times 10^{-10}\,\mu_B $, while LAMPF $\mu_{\nu_\mu} < 7.4 \times 10^{-10}\,\mu_B$~\cite{Allen:1992qe} and BNL-E734 $\mu_{\nu_\mu} < 8.5 \times 10^{-10}\,\mu_B $~\cite{Ahrens:1990fp}. Bounds on the tau neutrino magnetic moment are significantly weaker due to the experimental difficulty in producing intense $\nu_\tau$ beams. The best limits for $\mu_{\nu_\tau}$ are of the order $\mu_{\nu_\tau} \lesssim (4-5) \times 10^{-7}\,\mu_B$~\cite{Cooper-Sarkar:1991vsl}.

Astrophysical constraints are obtained from stars of varying mass and evolutionary stage. The flux of neutrinos produced in nuclear processes can be directly detected in neutrino and dark-matter experiments~\cite{Super-Kamiokande:2004wqk,Borexino:2017fbd,XMASS:2020zke,XENON:2022ltv,LZ:2022lsv,AtzoriCorona:2022jeb,Khan:2022bel,A:2022acy,PandaX:2022ood,LZ:2023poo,DeRomeri:2026tzh}, reaching down to $\mu_{\nu} \lesssim 6.4 \times 10^{-12}\,\mu_B$. More stringent constraints are obtained with hot and degenerate stars, in which plasmon decays---allowed by the dispersion relations of photons in such a dense environment---can abundantly contribute to energy loss~\cite{Raffelt:1989xu,Haft:1993jt,Raffelt:1996wa,Heger:2008er,Althaus:2010pi,Viaux:2013hca,MillerBertolami:2014oki,Corsico:2014mpa}. The bound from the white dwarf luminosity function is $\mu_\nu\lesssim 5.0\times 10^{-12}\mu_B$, and applies to all neutrino flavors~\cite{MillerBertolami:2014oki}. Similarly, strong bounds can be obtained by studying the luminosity at the tip of the red-giant branch. Recent bounds from NGC~4258 are extremely competitive, $\mu_\nu\lesssim 1.5\times 10^{-12}\mu_B$~\cite{Capozzi:2020cbu}; one could obtain even more stringent bounds from $\omega$-Centauri, though for such values of $\mu_\nu$ the energy loss is comparable to the uncertainty over the standard energy losses due to weak interaction, making it increasingly difficult to robustly exclude smaller dipole moments.

Finally, constraints arise from cosmological production. Right-handed states would increase the effective number of relativistic degrees of freedom, $N_{\mathrm{eff}}$~\cite{Morgan:1981zy,Fukugita:1987uy,Elmfors:1997tt,Ayala:1999xn}. These bounds are very sensitive to the cosmological history; using current BBN and CMB data, the authors of Ref.~\cite{Carenza:2022ngg} find $\mu_\nu\lesssim2.6\times 10^{-12}\mu_B$ assuming a reheating temperature $T_{\rm rh}=100\,\rm GeV$, which are raised to $\mu_\nu\lesssim1.6\times 10^{-11}\mu_B$ assuming $T_{\rm rh}=100\,\rm MeV$ (see also Refs.~\cite{Li:2022dkc,Grohs:2023xwa}).

\section{Conclusions and outlook\label{sec:conclusions}}

If neutrinos have dipole moments $\mu_\nu$, right-handed states with energy $E_\nu\simeq 100\,\rm MeV$ can be copiously produced in the core of a SN, mainly by spin-flip scattering on charged particles. Subsequently, these sterile neutrinos can reconvert back into active states in Galactic magnetic fields, yielding a high-energy flux that would have been detected by IMB and Kamiokande-II from SN~1987A. Our main result is the reassessment of this bound, which was only estimated in order of magnitude in the 1980s in Refs.~\cite{Lattimer:1988mf,Barbieri:1988nh,Notzold:1988kz}. We find a constraint $\mu_{\nu_e}\lesssim 4.8\times 10^{-13}\mu_B$, surpassing the traditional cooling bound $\mu_{\nu_e}\lesssim 3.3 \times 10^{-12}\mu_B$ by a factor of 7. Hence, the core energy loss through emission of $\nu_R$ is at most $\mathcal{O}(1\%)$ of the total binding energy. We conclude that, reexamining the production rates and building upon advances in SN simulations, as well as in the description of the Milky Way magnetic field, the most conservative constraints are a factor of 2 better than the rough bounds first obtained in Ref.~\cite{Barbieri:1988nh}, $\mu_\nu\lesssim 10^{-12}$, using this observable. Our bounds also largely beat cosmological ($\mu_{\nu_e}\lesssim 2.6 \times 10^{-12}\mu_B$, with $T_{\rm rh}=100\,\rm GeV$) and laboratory ($\mu_{\nu_e}\lesssim 2.9\times 10^{-11}\mu_B$ from GEMMA~\cite{Beda:2012zz}) constraints.

Akin to the bounds on Majoron decays that two of us have proposed in Ref.~\cite{Fiorillo:2022cdq}, these constraints are based on the non-detection of high-energy neutrinos in 1987, as confirmed by private communication therein. By its very nature, this observable is much more robust than the duration of the neutrino burst usually adopted for the cooling bounds: on the one hand, it does not rely on the sparse statistics of the neutrino flux from SN~1987A. On the other hand, and most importantly, the constrained right-handed emission is so small that it has perturbative effects on the SN evolution itself, and is insensitive to possible effects modifying the cooling speed of the nascent PNS or its putative early collapse to a black hole.

The next Galactic SN will likely be a testbed for the existence of dipole moments. Assuming a SN at a distance of $10\,\rm kpc$ and rescaling the results obtained for IMB, we find that Hyper-Kamiokande will be sensitive to dipole moments in the $10^{-14}-10^{-13}\mu_B$ range. A more powerful reach will be obtained through a dedicated analysis that accounts for the different efficiency and energy resolution of Hyper-Kamiokande~\cite{Hyper-Kamiokande:2018ofw}; and that accounts for the standard SN signal, since the much larger statistics will allow it. Timing information will unambiguously discriminate between the existence of dipole moments and the existence of MeV-GeV FIPs mimicking this phenomenon in the energy spectra~\cite{Fiorillo:2022cdq,Akita:2022etk}, as the high-energy neutrino flux obtained from the latter would reach Earth after accumulating a delay~\cite{Telalovic:2024cot}, though it will be more challenging to distinguish the two cases if the decaying FIPs are lighter.

\section*{Acknowledgments}

MLR and EV are supported by the Italian Ministero dell'Università e della Ricerca through the FIS 2 project FIS-2023-01577 (DD n. 23314 10-12-2024, CUP C53C24001460001). DFGF, MLR, and EV acknowledge support from Istituto Nazionale di Fisica Nucleare (INFN) through the Theoretical Astroparticle Physics (TAsP) project. EV is also supported by the Departments of Excellence grant 2023--2027 ``Quantum Frontier''. Funded by the European Union (ERC, NEPTUNE, No. 101301124). Views and opinions expressed are however those of the author(s) only and do not necessarily reflect those of the European Union or the European Research Council. Neither the European Union nor the granting authority can be held responsible for them. The ERC funding acknowledged here supports the contribution of D.F.G.F. to this work.

\appendix
\section{Derivation of the polarization tensor} \label{app:polarization}
In this Appendix, we detail the derivation of the electromagnetic response in the PNS plasma, characterised by the photon polarization tensor $\Pi^{\mu\nu}(\omega,\mathbf{q})$, where $Q^\mu=(\omega,\mathbf{\mathbf{q}})$ is the four-momentum carried by the virtual photon. In linear response, the current induced in the plasma by an electromagnetic field $A^\nu$ is (see e.g. Ref.~\cite{kirzhnits1987general})
\begin{equation}
J^\mu_{\rm med}(Q)=\,\Pi^{\mu}{}_{\nu}(Q)\,A^\nu(Q).
\end{equation}
By Kubo's formula \cite{Kubo:1957mj}, $\Pi^\mu{}_\nu$ is given by the retarded correlator of the electromagnetic current $J^\mu=e\bar\psi\gamma^\mu\psi$,
\begin{equation}
\Pi^{\mu}{}_{\nu}(Q)= i\int d^4X\,e^{iQ\cdot X}\,\Theta(X^0)\,\langle[J_\nu(0),J^\mu(X)]\rangle ,
\end{equation}
where the Heaviside function expresses causality and enforces the vanishing of the polarization tensor for negative times. 
Retarded correlators cannot be evaluated directly with Feynman diagrams. We use instead the fact that the real part of the retarded function coincides with that of the time-ordered one, namely $\mathrm{Re}[\Pi_{\nu}^{\mu}]=\mathrm{Re}[\tilde\Pi_{\nu}^{\mu}]$, while their imaginary parts differ only in the position of the poles in $\omega$. Causality requires the retarded function to be analytic in the upper half of the plane, which is ensured by the prescription $\omega\to\omega+i\epsilon$.
We therefore take as the working definition of the polarization tensor the time-ordered correlator,
\begin{equation}
\tilde{\Pi}^{\mu}{}_{\nu}(Q)=- i\int d^4X\,e^{iQ\cdot X}\,\langle \mathcal{T}\,J_\nu(0)J^\mu(X)\rangle ,
\label{eq:Pi_timeordered}
\end{equation}
which can be computed with the ordinary Feynman rules in the medium, and we set $\omega\to\omega+i\epsilon$ at the end. The two current operators correspond to two insertions of the current operator with momentum $Q$, which are connected by the time ordered fermion propagator
\begin{equation}
\tilde{\Pi}^{\mu}{}_{\nu}(Q)=- i e^2\int\frac{d^4P}{(2\pi)^4}\,{\rm Tr}\big[\gamma^\mu S_F(P+Q)\,\gamma_\nu S_F(P)\big],
\label{eq:Pi_loop}
\end{equation}
where $S_F$ is the Feynman propagator in the medium 
\begin{equation}
S_F(P)=(\slashed P+m)\Big[\frac{1}{P^2-m^2+i0}+2\pi i \delta(P^2-m^2)\big(\Theta(P^0)\,n_{\mathbf p}+\Theta(-P^0)\,\bar n_{\mathbf p}\big)\Big]
\end{equation}
with $n_{\mathbf p}$ for particles  and $\bar n_{\mathbf p}$ for antiparticles. Defining $f_{\mathbf{p}}= n_{\mathbf{p} }+\bar n_{\mathbf{p}}$, one finds (see also Refs.~\cite{Altherr:1992mf,Braaten:1993jw,Raffelt:1996wa})
\begin{equation}
\Pi^{\mu}{}_{\nu}(Q)= 4 e^2\int\frac{d^3\mathbf{p}}{(2\pi)^3\,2E_\mathbf{p}}\,f_\mathbf{p}\sum_{\zeta=\pm}
\frac{2P^\mu P_\nu+\zeta\big(P^\mu Q_\nu+P_\nu Q^\mu-\delta^\mu_\nu\,P\!\cdot\!Q\big)}{\big(E_\mathbf{p}+\zeta(\omega+i0)\big)^2-E^2_{\mathbf p+\zeta\mathbf q}}.
\label{eq:Pi_final}
\end{equation}
Such structure has a simple physical interpretation, namely, the forward scattering of a photon with arbitrary $Q^\mu$ on all charged particles and antiparticles constituting the plasma.
Starting from $\Pi^{\mu}{}_{\nu}$, the longitudinal and transverse components of the polarization tensor read $\Pi ^{00}=\Pi_L$ and $\Pi^a_a=\Pi_T$, respectively. 
The imaginary parts of $\Pi_{L,T}$ measure the fluctuations of the medium, while the real parts in the denominators describe screening. All the dependence on temperature, density and degeneracy of the plasma enters through $\Pi_{L,T}$.
 If we call $\mathcal{X}=\cos\theta$ the cosine of the angle between $\mathbf{q}$ and $\mathbf{p}$, we finally find
\begin{equation}
\Pi_L(Q)=\frac{e^2}{2\pi^2}\int_0^\infty\frac{p^2f_\mathbf{p}\,dp}{E_\mathbf{p}}\int_{-1}^{1}d\mathcal{X}\sum_{\zeta=\pm}
\frac{2E_\mathbf{p}^2+\zeta(E_\mathbf{p}\omega+pq\mathcal{X})}{(\omega+i0)^2-q^2+\zeta\left[2E_\mathbf{p}(\omega+i0)-2pq\mathcal{X}\right]}.
\label{eq:Pi_longitudinal}
\end{equation}
For the transverse component, instead, we may use the Ward identity $Q_\mu\Pi^{\mu\nu}=0$ to obtain
\begin{equation}
\Pi_T(Q)=\frac12\left[\Pi^{\mu}{}_{\mu}+\frac{Q^2}{q^2}\,\Pi_L\right],
\end{equation}
where
\begin{equation}
\Pi^{\mu}{}_{\mu}(Q)=\frac{e^2}{\pi^2}\int_0^\infty\frac{p^2f_\mathbf{p}\,dp}{E_\mathbf{p}}\int_{-1}^{1}d\mathcal{X}\sum_{\zeta=\pm}
\frac{m^2-\zeta(E_\mathbf{p}\omega-pq\mathcal{X})}{(\omega+i0)^2-q^2+\zeta\left[2E_\mathbf{p}(\omega+i0)-2pq\mathcal{X}\right]} .
\end{equation}

We now compute the contribution to the polarization tensor from relativistic electrons.
For the longitudinal component, we start from Eq.~\eqref{eq:Pi_longitudinal} and take the limit of
massless electrons. This limit is safe, with one exception: when $\omega$ is
very close to $q$, the small deviation from the light cone induced by the electron mass may decide whether or not the virtual photon can be absorbed on shell. Since this range covers only a limited portion of the parameter space, we do not worry about it and simply take the massless electron limit. For the same reason, in the denominator we neglect terms of order
$\omega^2, q^2 \ll p\omega, pq$, so that
\begin{equation}
  \Pi_L(\omega,\mathbf{q}) = \frac{e^2}{2\pi^2}
  \int_0^{+\infty} \frac{p^2 f_\mathbf{p}\,\mathrm{d}p}{E_\mathbf{p}}
  \int_{-1}^{+1} \mathrm{d}\mathcal{X}
  \left(
    \frac{\omega + q\mathcal{X}}{\omega - q\mathcal{X} + i0}
    - \frac{\omega^2 - q^2}{(\omega - q\mathcal{X} + i0)^2}
  \right).
  \label{eq:Pi_longitudinal_e}
\end{equation}
The integral can now be solved explicitly, giving
\begin{equation}
  \Pi_L(\omega,\mathbf{q}) = \frac{e^2}{\pi^2}
  \left(\frac{\hat{\mu}_e^2}{2} + \frac{\pi^2 T^2}{6}\right)
  \left[
    \frac{\omega}{q}\,
    \log\!\left(\frac{\omega + q + i0}{\omega - q + i0}\right) - 2
  \right].
  \label{eq:PiL-explicit}
\end{equation}
In the limit $\omega \gg q$ this tends to
$\Pi_L \to \omega_{\mathrm{pl},e}^2\, q^2/\omega^2$, where
$\omega_{\mathrm{pl},e}$ is the electronic plasma frequency,
so we may replace
\begin{equation}
  \Pi_L(\omega,q) = \frac{3\,\omega_{\mathrm{pl},e}^2}{2}
  \left[
    \frac{\omega}{q}\,
    \log\!\left(\frac{\omega + q + i0}{\omega - q + i0}\right) - 2
  \right].
  \label{eq:PiL-plasma}
\end{equation}
The polarization tensor is real for $|\omega| > |q|$, since superluminal photons cannot be emitted by relativistic electrons via Cherenkov radiation. Due to the logarithm, however, for $|\omega| < |q|$ it acquires an imaginary part
\begin{equation}
  \operatorname{Im}\left[\Pi_L(\omega,q)\right]
  = -\frac{e^2\,\omega}{\pi q}
  \left(\frac{\hat{\mu}_e^2}{2} + \frac{\pi^2 T^2}{6}\right)
  \Theta\!\left(q^2 - \omega^2\right).
  \label{eq:Im_Pi_longitudinal_e}
\end{equation}
The trace $\Pi^\mu{}_\mu$ is even easier to determine, since the logarithmic singularity cancels out completely in the limit we have chosen,
\begin{equation}
  \Pi^\mu{}_\mu = -\frac{2e^2}{\pi^2}
  \left(\frac{\hat{\mu}_e^2}{2} + \frac{\pi^2 T^2}{6}\right)
  \label{eq:Pi_trace}
\end{equation}
and thus it has no imaginary part. Thus, the transverse component reads
\begin{equation}
  \Pi_T(\omega,q) = \frac{3\,\omega_{\mathrm{pl},e}^2}{4q^3}
  \left[
    \omega\left(\omega^2 - q^2\right)
    \log\!\left(\frac{\omega + q + i0}{\omega - q + i0}\right)
    - 2q\omega^2
  \right].
  \label{eq:Pi_transverse_e}
\end{equation}

For protons, we start from Eq.~\eqref{eq:Pi_longitudinal} and take the non-relativistic limit $E_\mathbf{p}\to m_N$, leading to
\begin{equation}
  \Pi_L(\omega,\mathbf{q}) = \frac{e^2}{2\pi^2}
  \int_0^{+\infty} \frac{p^2 f_\mathbf{p}\,\mathrm{d}p}{m_N}
  \int_{-1}^{+1} \mathrm{d}\mathcal{X}\,
  \frac{q^2}{(\omega - vq\mathcal{X} + i0)^2},
  \label{eq:Pi_longitudinal_p}
\end{equation}
where $v = p/m_N$. The same expression could be obtained more easily from the Vlasov equation. After performing the angular integration we find
\begin{equation}
  \Pi_L(\omega,\mathbf{q}) = \frac{e^2}{\pi^2}
  \int_0^{+\infty} \frac{p^2 f_\mathbf{p}\,\mathrm{d}p}{m_N}\,
  \frac{q^2}{(\omega - vq + i0)(\omega + vq + i0)}.
  \label{eq:PiL-protons-2}
\end{equation}
For a generic degree of degeneracy, the integral must be performed numerically, with
\begin{equation}
  f_\mathbf{p} = \frac{1}{e^{(m_N-\mu_p+p^2/2m_N)/T}+1}.
  \label{eq:fp-protons}
\end{equation}
The imaginary part, however, can be expressed analytically. After rewriting
\begin{equation}
  \Pi_L(\omega,\mathbf{q}) = \frac{e^2 q}{2\pi^2}
  \int_0^{+\infty} p f_\mathbf{p}\,\mathrm{d}p
  \left[\frac{1}{\omega - vq + i0} - \frac{1}{\omega + vq + i0}\right],
  \label{eq:PiL-protons-3}
\end{equation}
we find
\begin{equation}
  \operatorname{Im}\left[\Pi_L(\omega, \mathbf{q})\right]
  = -\frac{e^2 m_N^2\,\omega}{2\pi q}\, f_{m_N|\omega|/q}.
  \label{eq:Im_Pi:longitudinal_p}
\end{equation}

\section{Emissivity of right-handed neutrinos}\label{app:emissivity}
From the Lagrangian of Eq.~\eqref{eq:lagrangian} the number of $\nu_R$ emitted with momentum $\mathbf k$ per unit volume and time is
\begin{equation}
\dot{\mathcal{N}}_{\nu_R,\mathbf{k}}
=\frac{\mu_\nu^2}{E_\mathbf{k}} \int \frac{d^3\mathbf{p}}{(2\pi)^3\,2E_\mathbf{p}}\,
n^{\nu_L}_{\mathbf{p}}\, (P\cdot K)(P_\nu+K_\nu)(P_\beta+K_\beta)
\int d^4X\, \langle A^\nu(0)A^\beta(X)\rangle\, e^{i(K-P)\cdot X},
\label{eq:emissivity_correlator}
\end{equation}
where $P^\mu$ and $K^\mu$ are the four-momenta of $\nu_L$ and $\nu_R$ (massless,
$P^2=K^2=0$), spins are summed and $Q^\mu=K^\mu-P^\mu=(\omega,\mathbf q)$ is the
four-momentum extracted from the medium. 

The Fourier transform of the correlator
is the virtual occupation number of the photon field. By applying the fluctuation-dissipation theorem, the spacetime correlator inside the integral can be directly related to the retarded Green's function. Indeed, the two-point correlation function is defined as the Wightman function $G^{<\mu\nu}(X) = \langle A^\mu(0)A^\nu(X)\rangle$. The retarded photon propagator is defined as $D_R^{\mu\nu}(X) = -i\theta(t) \langle [A^\mu(X), A^\nu(0)] \rangle$. By taking the Fourier transform over the spacetime coordinates with momentum transfer $Q = K-P$, these two quantities are related as
\begin{equation}
\int d^4X \langle A^\mu(0)A^\nu(X)\rangle e^{iQ\cdot X} = -\frac{2}{e^{\beta\omega}-1} \text{Im}\, D_R^{\mu\nu}(Q).
\label{eq:corrtoprop}
\end{equation}
In turn, the retarded propagator is found by
\begin{equation}
  A^\mu(X)=\int d^4Y\,D_R^{\mu\nu}(X-Y)\,J_{\mathrm{ext},\nu}(Y).
\end{equation}
In the medium, Maxwell's equations read
$\partial_\alpha F^{\alpha\mu}=J^\mu_{\rm ext}+J^\mu_{\rm med}$.  In Coulomb gauge, $\nabla\!\cdot\!\mathbf A=0$, and for
$Q^\mu=(\omega,0,0,q)$ we find in Fourier space
\begin{equation}
  A^0=\frac{J^0}{q^2-\Pi_L(Q)},\qquad A^a=-\frac{J^a}{Q^2+\Pi_T(Q)},
\end{equation}
where $a=x,y$ labels the two transverse modes and $\Pi_L=\Pi^{00}$,  $\Pi_T=\Pi^a{}_a$. Therefore, the retarded photon propagator reads in Fourier space
\begin{equation}
  (D_R)^0{}_0=\frac{1}{q^2-\Pi_L(Q)},\qquad
  (D_R)^i{}_j=-\frac{1}{Q^2+\Pi_T(Q)}\left(\delta_{ij}-\frac{q_iq_j}{q^2}\right).
  \label{eq:D-coulomb}
\end{equation}
Strictly speaking, the retarded longitudinal propagator in Coulomb gauge does not
coincide with the commutator, because $A^0$ is not a dynamical degree of freedom
and its commutator vanishes in the absence of a medium; one should subtract the
vacuum value of $(D_R)^0{}_0$. However, the imaginary part is unaffected by this
procedure, and only the imaginary part enters Eq.~\eqref{eq:corrtoprop}. We thus find the right-handed neutrino emissivity
\begin{equation}
\begin{aligned}
\dot{\mathcal{N}}_{\nu_R,\mathbf{k}}
&=\frac{2\mu_\nu^2}{E_\mathbf{k}}
\int\frac{d^3\mathbf p}{(2\pi)^3\,2E_\mathbf{p}}\,
n^{\nu_L}_{\mathbf p}\,
\frac{P\!\cdot\!K}{e^{\beta(k-p)}-1}
\\
&
\left[
-\frac{(p+k)^2\,\mathrm{Im}\,\Pi_L(K-P)}
{\big|\,|\mathbf k-\mathbf p|^2-\Pi_L(K-P)\big|^2}
+\frac{4|\mathbf k\times\mathbf p|^2\,\mathrm{Im}\,\Pi_T(K-P)}
{|\mathbf k-\mathbf p|^2
\big|(K-P)^2+\Pi_T(K-P)\big|^2}
\right].
\end{aligned}
\label{eq:emissivity-general}
\end{equation}
So far we have made no approximation beyond linear response. The whole problem now
reduces to determining the current that the medium produces in response to an
external field, which enters only through $\Pi_L$ and $\Pi_T$.

We can now determine the contributions to the emissivity arising from virtual fluctuations sourced by electrons. By substituting the expressions obtained in  Appendix~\ref{app:polarization}, we find
\begin{equation}
\begin{aligned}
\dot{\mathcal{N}}_{\nu_R,\mathbf{k}}^{\rm electrons}
=&
\frac{3\pi\mu_\nu^2\omega_{\rm pl,e}^2}{k}
\int \frac{d^3\mathbf{p}}{(2\pi)^3\,2p}\,
n_{\mathbf{p}}^{\nu_L}
\frac{pk(1-\cos\theta)}{e^{\beta(k-p)}-1}
\\
&
\frac{(k-p)}
{|\mathbf{p}-\mathbf{k}|}
\left[
\frac{(p+k)^2}
{\left|\,|\mathbf{k}-\mathbf{p}|^2
-\Pi_L(K-P)\right|^2}
\right.\left.
+\frac{4p^3k^3\sin^2\theta(1-\cos\theta)}
{|\mathbf{k}-\mathbf{p}|^4
\left|(K-P)^2+\Pi_T(K-P)\right|^2}
\right],
\end{aligned}
\end{equation}
where $\theta$ denotes the angle between $\mathbf{p}$ and $\mathbf{k}$, and we have assumed the neutrinos to be massless. By introducing the new integration variables $\omega=k-p$ and $q=|\mathbf{k}-\mathbf{p}|$, the expression can be rewritten as

\begin{equation}
\begin{aligned}
\dot{\mathcal{N}}_{\nu_R,\mathbf{k}}^{\rm electrons}
=&
\frac{3\mu_\nu^2\omega_{\rm pl,e}^2}{16\pi k^2}
\int_{-\infty}^{k} d\omega\,
n_{k-\omega}^{\nu_L}
\frac{\omega}{e^{\beta\omega}-1}
\\
&
\int_{|\omega|}^{2k-\omega} dq\,
(q^2-\omega^2)
\left[
\frac{(2k-\omega)^2}
{|q^2-\Pi_L|^2}
\right.\left.
+ \ \frac{(q^2-\omega^2)^2
\left[(2k-\omega)^2-q^2\right]}
{2q^4|q^2-\omega^2-\Pi_T|^2}
\right].
\end{aligned}
\end{equation}

Since $\omega$ and $q$ are in no definite hierarchy, no special simplification is
available for $\Pi_{L,T}$. However, for typical energies much larger than the
plasma frequency we can obtain a logarithmic approximation that captures the
correct magnitude. If $\Pi_{L,T}$ are neglected in the denominators, the integral
over $q$ is elementary and gives $4(k-\omega)^2/\omega$ for $\omega>0$ and
$-4k^2/\omega$ for $\omega<0$. Changing variable $\omega\to-\omega$ in the
negative region,
\begin{equation}
\dot{\mathcal{N}}_{\nu_R,\mathbf{k}}^{\rm electrons}=\frac{3\mu_\nu^2\omega_{\rm pl,e}^2}{4\pi k^2}
\left[\int_0^{\infty}d\omega\,\frac{n_{k+\omega}\,k^2}{1-e^{-\beta\omega}}
+\int_0^{k}d\omega\,\frac{n_{k-\omega}\,(k-\omega)^2}{e^{\beta\omega}-1}\right].
\label{eq:emissivity-electrons-nomedium}
\end{equation}
The first integral describes processes in which the left-handed neutrino loses
energy to the medium, the second one the opposite case. Neglecting the medium
response makes both integrals diverge logarithmically for $\omega\to0$. Since the
medium properties enter only logarithmically, it is enough to estimate the scale at
which the approximation breaks down and use it as a lower cut. The electronic
$\Pi_L$ is of order $\omega_{\rm pl,e}^2$ for $q\gtrsim\omega$ (the region
$q\simeq|\omega|$ is suppressed by the factor $q^2-\omega^2$), the transverse part behaves similarly, and the protons give $\Pi_L/q^2\lesssim\omega_{\rm pl,p}^2/\omega^2$.
Hence the medium response can be neglected for $|\omega|\gg\omega_{\rm pl}$, where $\omega_{\rm pl}$ is the total plasma frequency, dominated by electrons.

For non-relativistic protons we can neglect the transverse photons, since
non-relativistic nucleons primarily emit electric fields. Moreover,
$\mathrm{Im}\,(\Pi_L)$ is exponentially suppressed when the phase velocity $|\omega|/q$ exceeds the thermal proton velocity $v_N\ll1$, so we can take
$|\omega|\ll q$ everywhere. Then $\omega\ll T$ and
$e^{\beta\omega}-1\simeq\omega/T$, $P\!\cdot\!K\simeq q^2/2$, $p+k\simeq2k$.   After expressing the integration in
terms of $\omega$ and $q$,
\begin{equation}
\dot{\mathcal{N}}_{\nu_R,\mathbf{k}}^{\rm protons}=\frac{\mu_\nu^2\,T\,n_k}{2\pi^2}\int_0^{2k}q^3dq
\left[-\int_{-\infty}^{+\infty}\frac{d\omega}{\omega}\,
\frac{\mathrm{Im}\,(\Pi_L(Q))}{|q^2-\Pi_L(Q)|^2}\right].
\label{eq:proton-omega}
\end{equation}
In the numerator only the proton contribution to $\Pi_L$ is included, while the
denominator contains also the electrons. In the range $\omega\sim v_Nq\ll q$ where
the integral is performed, the response is dominated by protons, which justifies
using the proton $\Pi_L$ everywhere. We now note that
\begin{equation}
  \frac{\mathrm{Im}\,(\Pi_L)}{|q^2-\Pi_L|^2}
  =\mathrm{Im}\left(\frac{\Pi_L}{q^2\,(q^2-\Pi_L)}\right),
\end{equation}
where the subtraction makes the function in brackets vanish for $\omega\to\pm\infty$. It has no poles in the upper half-plane, since the medium is
in equilibrium and hence stable, so the Kramers-Kronig relation enforces the equality
\begin{equation}
  \int_{-\infty}^{+\infty}\frac{d\omega}{\omega}\,
  \mathrm{Im}\left(\frac{\Pi_L(Q)}{q^2(q^2-\Pi_L(Q))}\right)
  =\frac{\pi}{q^2}\,\mathrm{Re}\left( \frac{\Pi_L(0,q)}{q^2-\Pi_L(0,q)}\right).
\end{equation}
The use of the Kramers-Kronig relation to simplify the calculation of emission rates in the classical, zero-recoil limit was first used, to our knowledge, by Raffelt in the context of Primakoff emission of axions~\cite{Raffelt:1987np}.
We can now use the fact that at $\omega=0$ the polarization
$\Pi_L(0,q)=-k_{D,p}^2$ is purely real, with $k_{D,p}$ the proton Debye scale. Inserting in Eq.~\eqref{eq:proton-omega} and performing the integral over $q$,
\begin{equation}
\dot{\mathcal{N}}_{\nu_R,\mathbf{k}}^{\rm protons}
=\frac{\mu_\nu^2\,T\,n_k}{4\pi}\,k_{D,p}^2\,
\log\!\left(1+\frac{4k^2}{k_{D,p}^2}\right).
\label{eq:emissivity-protons}
\end{equation}
which is Eq.~\eqref{eq:proton_emissivity}. 

Parametrically, when we integrate over $d^3\bk \,\omega/(2\pi)^3$, we find a cooling rate that scales as $\mathcal{Q}_p\sim \mu_\nu^2 T k_{D,p}^2 \hat{\mu}_{\nu_e}^4$ for the degenerate $\nu_e$, and $\mathcal{Q}_p\sim \mu_\nu^2 T^5 k_{D,p}^2$ for the nondegenerate species. This gives us an opportunity to also confirm that plasmon decay, that we have neglected so far, is indeed largely subdominant. For plasmon decay, Eq.~(6.92) of Ref.~\cite{Raffelt:1996wa} shows that the cooling rate is parametrically of the order $\mathcal{Q}_\gamma\sim \mu_\nu^2 T^3 \omega_{\rm pl}^4$. Since $\omega_{\rm pl}^4\ll T^2 k_{D,p}^2$, we find explicitly that it is a subdominant correction in comparison with $\mathcal{Q}_p$ even for the nondegenerate $\nu_\tau$.

\bibliographystyle{bibi}
\bibliography{references.bib}

\end{document}